\documentclass[11pt,a4paper]{article}
\usepackage[utf8]{inputenc}
\usepackage[OT1]{fontenc}
\usepackage{amsmath}
\usepackage{amsfonts}
\usepackage{amssymb}
\usepackage{amsthm}
\usepackage{bbm}
\usepackage{enumerate}
\usepackage{graphbox,graphicx}
\usepackage{wrapfig}
\graphicspath{ {./photos/} }
\usepackage{xcolor}
\usepackage[left=2.9cm, right=2.90cm,top=3.10cm, bottom=3.10cm]{geometry}
\usepackage{hyperref}

\makeatother
\hypersetup{
    colorlinks=true,
    citecolor=red,
    filecolor=black,
    linkcolor=black,
    urlcolor=blue
}
\usepackage{csquotes}

\DeclareFontFamily{OT1}{rsfs}{}
\DeclareFontShape{OT1}{rsfs}{m}{n}{ <-7> rsfs5 <7-10> rsfs7 <10-> rsfs10}{}
\DeclareMathAlphabet{\mycal}{OT1}{rsfs}{m}{n}
\DeclareMathAlphabet\mathbfcal{OMS}{cmsy}{b}{n}
\theoremstyle{definition}
\newtheorem{df}{Definition}[section]
\theoremstyle{plain}
\newtheorem{sthm}{Proposition}[section]

\newtheorem{thm}{Theorem}[section]

\newcommand{\mcH}{\mycal H}
\newcommand{\mcL}{{\mycal L}}

\newcommand{\Lie}{{\mathcal{L}}}
\newcommand{\myLie}{\mathbf{L}}

\newcommand{\mb}[1]{\boldsymbol{#1}}
\newcommand{\LKmom}{\mathcal{P}_{d S}}
\newcommand{\LKbst}{\mathcal{L}_{d S}}
\newcommand{\LKmcon}{p}
\newcommand{\LKbcon}{l}
\newcommand{\zspaceD}{ {\mathring D}}
\newcommand{\wtx}{n}

\newcommand{\EdS}{\mathit{E}_{H}}
\newcommand{\JdS}{{J}}
\newcommand{\KilRx}{\stackrel{\scriptscriptstyle (x)}{\KilR}}
\newcommand{\KilRy}{\stackrel{\scriptscriptstyle (y)}{\KilR}}
\newcommand{\KilRz}{\stackrel{\scriptscriptstyle (z)}{\KilR}}
\newcommand{\JdSx}{\stackrel{\scriptscriptstyle(x)}{\JdS}\! \! \!{}_{H}}
\newcommand{\JdSy}{\stackrel{\scriptscriptstyle(y)}{\JdS}\! \! \!{}_{H}}
\newcommand{\JdSz}{\stackrel{\scriptscriptstyle(z)}{\JdS}\! \! \!{}_{H}}

\newcommand{\chitrho}{\chi_{\scriptscriptstyle H}}
\newcommand{\chitrhobar}{\overline{\chi}_{\scriptscriptstyle H}}

\newcommand{\mcLHop}{\stackrel{H}{\mcL}}
\newcommand{\KilT}{\mathcal{T}}
\newcommand{\KilR}{\mathcal{R}}
\newcommand{\Hor}{\mathcal{H}_{\mathcal{C}}}

\title{Hopfion-like solutions in de Sitter spacetime}
\author{Adam Grzela, Jacek Jezierski, Tomasz Smo\l ka\thanks{Corresponding author:t.smolka@uw.edu.pl}\\
Department of Mathematical Methods in Physics\\
 Faculty of Physics, University of Warsaw \\
Pasteura 5, 02-093, Warsaw, Poland}
\date{March 20, 2024}
\begin{document}
	\numberwithin{equation}{section}
    \maketitle
    \begin{abstract}
    	We construct electromagnetic field with non-trivial topological properties on de Sitter background. The field is closely related with Hopf fibration. We analyze energy, angular momentum and topological charges for this solution.
        The paper is a generalization of CQG \textbf{35} (2018), no. 24, 245010 to de Sitter spacetime.
    \end{abstract}
    \section{Introduction}
    		
    		Hopfion is a ‘solitonary’ solution of the spin-N field (including electromagnetic and gravitational) with a rich
    		topological structure related to Hopf fibration. The Hopf fibration is the simplest non-trivial fibration of a three-dimensional sphere. The Hopf map, associated with the fibration, is a surjective map sending
    		great circles on $\mathbb{S}^3$ to points on $\mathbb{S}^2$. These circles weave nested toroidal surfaces, and each is linked with every other circle exactly once, creating the characteristic structure of Hopf fibration (see section \ref{sec:TopoProperties}). The structure of hopfions is visible on the integration curves of the vector field, for example see \cite{irvine2008linked}. The closed and linked field lines of hopfions propagate with no intersections along the light cone. These exotic solutions arise in various fields of physics, such as electromagnetism \cite{ranada2004topological,irvine2008linked}, magnetohydrodynamics \cite{kamchatnov1982topological}, hadronic physics \cite{skyrme1962unified}, or Bose--Einstein condensate \cite{kawaguchi2008knots}.
    		
   			Recently, hopfions as topologically non-trivial magnetic configurations have garnered significant attention \cite{MagnetHopfions1Saji2023}. The scattering of spin waves by magnetic hopfions has been analyzed using a micromagnetic approach. The spin waves propagating along the hopfion symmetry axis are deflected by the magnetic texture, acting as a convergent or divergent lens \cite{MagnetHopfions2Sallermann2023}. Spin waves propagating along the plane perpendicular to the symmetry axis respond with a skew scattering and a closely related Aharonov-Bohm effect.
    			
   			In last years, Crisan, Godinho, and Vancea \cite{cricsan2021gravitoelectromagnetic} proposed a new construction of knotted solutions for weak gravitational fields, analogous to the Ra\~nada hopfion fields. The solutions have been obtained with the help of time-dependent gravitoelectromagnetic (GEM) equations. The analysis of the Landau-Lifshitz pseudo-tensor for the static fields provides insight into the energy-momentum distribution for the obtained gravito-electromagnetic knot fields.

    		The recent gravitational waves detections \cite{LIGO3rdrun2017} and the measurement of a positive cosmological constant \cite{planck_ade2014} lead to an urgent need of a thorough understanding of radiation in spacetimes with a positive cosmological constant $\Lambda$. The topological properties of electromagnetic and gravitational solutions are poorly understood, even in maximally symmetric spacetime with a positive cosmological constant, known as de Sitter spacetime. From a topological perspective, the simplest solution of Einstein’s equations with a positive cosmological constant is far more intriguing than Minkowski’s spacetime. De Sitter geometry can be seen as a Lorentz hyperboloid in five-dimensional Minkowski space with the topology represented as $\mathbb{R} \times \mathbb{S}^3$, indicating that it is spatially compact. De Sitter spacetime has future and past event horizons. The symmetry group is SO(4, 1), the de Sitter group. For more details, see section \ref{sec:deSitter} or classical reference \cite[Chapter 4]{griffiths2009exact}.  
    		
    		The aim of the paper is to present a topological analysis of a Hopfion-like solution on the de Sitter background. First, we obtain a Maxwell field on the de Sitter background with analogical properties to hopfions. Then, we attempt to find how the structure of Hopf fibration interferes with the topology of de Sitter spacetime. Our research contains investigations near the cosmological horizon.
    		
    		The paper is organized as follows:
    		In section \ref{sec:deSitter}, we briefly discuss the properties of de Sitter spacetime together with the coordinate systems that are most suitable in our context. In the following part (section \ref{sec:GenerHopf}), we use an original description of reduced data\footnote{The reduced data are discussed in appendix \ref{sec:ReducedDataDeSitter}.} and conformal covariance of Maxwell theory to find the generalization of hopfions in de Sitter spacetime. Section \ref{sec:TopoProperties} starts with a brief recall of Hopf fibration. Next, we pass to the topological analysis of generalized hopfions near the cosmological horizon. We identify the Hopf structure in the solution and construct an integral curve that reaches the horizon for finite values of parametrization. The analysis of classical charges, such as energy, angular momentum and their fluxes is done in section \ref{sec:KillingCharges}. Next, we pass to the study of topological charges in section \ref{sec:TopoCharges}. 
    		
    		The appendix \ref{sec:ReducedDataDeSitter} contains a description of the Maxwell field in terms of proposed reduced data. In particular, the relations between Maxwell’s two-form and reduced data are given. The description of reduced data is supported by a brief survey of the mathematical background (appendix \ref{sec:math_sup}). Next, section \ref{sec:ElCYK} contains a brief survey of conformal Yano--Killing two-forms and their applications in electrodynamics. We briefly describe how to construct reduced data for Pleba\'nski--Demia\'nski family of generalized black hole spacetimes. Appendix \ref{sec:KillingVF} contains the whole basis of Killing vactor fields for de Sitter spacetime. In appendix \ref{sec:UniqnessDesitter}, we briefly analyze ambiguities of explicitly conformally flat form of de Sitter.
    	
    	\subsection{Notation and conventions}

    		In the paper, we utilize geometric coordinate systems with $c = G = 1$ and abstract index notation. To make our analysis more convenient, we use the Einstein summation convention, where the summation depends on the type of indices (which are explained later on). We examine Maxwell's theory on a four-dimensional de Sitter background. We describe the de Sitter background in three coordinate systems:
    		\begin{itemize}
    			\item Conformally flat coordinates $(T,R,x^{A})$ with the metric in the form \eqref{eq:dSMetricConformal}
    			\item Stationary coordinates $(t,r,x^{A})$ for which the metric is given by \eqref{eq:deSitterStationary} 
    			\item Coordinates $(t,\rho,x^{A})$ with the metric \eqref{eq:CompactCoordMetric}  
    		\end{itemize}
    		In section \ref{sec:deSitter}, we define all three types of coordinates. $x^{A}$ denotes angular coordinates, which may or may not be explicitly given. Capital letters represent the two-dimensional space of indices of the axial coordinates on the unit sphere. For a distinguished angular coordinate representation, we denote $\theta$, $\varphi$ as the standard angles parameterizing a two-dimensional sphere with the two-metric $d \theta^2+\sin^2 \theta d \varphi $. Greek indices represent a four-dimensional set, while $x^{\mu}$ are not assigned to a specific choice of coordinates unless specified explicitly. Small Latin indices (excluding $t$ and $r$) represent spatial coordinates on the $\Sigma_{t}=\{t=\mathrm{const.}\}$ slice. Bold capital letters (like $\textbf{Z}, \textbf{E}$ or $\textbf{B}$) denote three-dimensional vector fields on $\{t=\mathrm{const.}\}$ slice. Unless necessary, we do not select a radial coordinate ($r$ or $\rho$). The symbol ',' represents the partial derivative $\partial$. The symbols $t$ and $r$ represent time and radius in static coordinates, respectively.
 
    		We will use $T_{...(\mu\nu)...}$ to represent the symmetric part and $T_{...[\mu\nu]...}$ to represent the antisymmetric part of tensor $T_{...\mu\nu...}$ concerning indices $\mu$ and $\nu$. Similar symbols will be used for more indices. The indications for measure (volume form), required when Stokes Theorem is applied, are described by \eqref{eq:StokesConvention}.
    		
    		$Y_{\ell}$ is a spherical harmonic with a specific degree that satisfies $\mathbf{\Delta} Y_{\ell}=-\ell (\ell+1) Y_{\ell}$. It can be expressed as a linear combination of spherical harmonics $Y_{\ell m}$ with any order $m$, where $m$ is within the range $\{-\ell,-\ell+1,...,\ell-1,\ell\}$. 
    		The formula defines the convention used for the Levi--Civita pseudo-tensor 
    		$\varepsilon_{\alpha \beta \gamma \delta}=\sqrt{-\det g_{\mu\nu}}\epsilon_{\alpha \beta \gamma \delta}$, where
    		\begin{equation}
    			\epsilon_{\alpha \beta \gamma \delta}=\left \{\begin{array}{l}
    				+1 \quad \mbox{if $\alpha \beta \gamma \delta$ is an even permutation of 0, 1, 2, 3}\, ,\\
    				-1 \quad \mbox{if $\alpha \beta \gamma \delta$ is an odd permutation of 0, 1, 2, 3}\, ,\\
    				\hspace{0.25 cm}0 \hspace{0.2 cm} \mbox{ in any other case} \, .
    			\end{array} \right.
    			\label{eq:4DLeviCivitaConvention}
    		\end{equation}

    	\subsection{Properties of de Sitter spacetime \label{sec:deSitter}}
    		De Sitter spacetime is a maximally symmetric solution of Einstein's equations with a positive cosmological constant. It can be represented as a Lorentz hyperboloid in five-dimensional Minkowski space. It is a spacetime of constant positive curvature with the structure of $\mathbb{R} \times \mathbb{S}^3$, which indicates that it is spatially compact. De Sitter spacetime has future and past event horizons, and its symmetry group is $SO(4, 1)$, also known as the de Sitter group. The de Sitter metric in stationary coordinates is commonly expressed as
    		\begin{equation}\label{eq:deSitterStationary}
    			g =   -f(r)^2dt^2 + \frac{dr^2}{f(r)^2}
    			+ r^2 \sigma_{A B} d x^{A} d x^{B} 
    			\, ,
    		\end{equation}
    		where $\sigma_{A B} d x^{A} d x^{B}$ is a standard unit sphere metric. If needed, we use an exact representation $x^{A}=(\theta, \varphi)$ with the two-metric
    		\begin{equation}
    			\sigma_{A B} d x^{A} d x^{B}=d\theta^2+\sin^2 \theta d\varphi^2 \, .
    			\label{eq:ExactTwoForm}	
    		\end{equation}
    		We have $f(r)=\sqrt{1-\alpha^2 r^2} \, , r \in \mathbb{R}_{+}\setminus \{\frac{1}{\alpha}\} $. We neglect the ordinary coordinate issues at the poles, where $\sin \theta=0$. The parameter $\alpha$, known as the Hubble constant, is defined by a positive cosmological constant as
    		\begin{equation}
    			\alpha=\sqrt{\frac{\Lambda}{3}} \, .
    		\end{equation} 
    		The null hypersurface
    		\begin{equation}
    			\Hor=\left\{(t,r,\theta,\varphi): r=\frac{1}{\alpha} \right\} \, ,
    			\label{eq:def_horizon}
    		\end{equation}
    		called a cosmological horizon, is an important concept in the de Sitter universe. The metric \eqref{eq:deSitterStationary} is degenerated on $\mathcal{H}_{\mathcal{C}}$.  The norm of the Killing vector $\partial_t$ vanishes on $\Hor$, so geometrically, it is a Killing horizon. We can distinguish the future event horizon on the null surface, where $r=\alpha^{-1}$ and $t=+\infty$. The past event horizon for the geodesic observer is located on the null surface, where $r=\alpha^{-1}$ and $t=-\infty$, at $r=0$. It is unique to the de Sitter universe that such horizons exist for geodesic observers, namely, the future and past event horizons. This is a direct consequence of the space-like character of the conformal infinities $\mathcal{I}^{+}$ and $\mathcal{I}^{-}$. However, such horizons do not occur for geodesic observers in Minkowski spacetime, because its conformal infinity is null. Only accelerating observers have the possibility of seeing such horizons in flat space, where their trajectories in the conformal diagram terminate at $\mathcal{I}^{\pm}$ rather than at time-like infinities $i^{\pm}$. Additionally, it is important to note that these cosmological horizons are observer-dependent.
    		
    		De Sitter spacetime is conformally flat. The metric can be transformed into an exact, conformally Minkowskian form
    		\begin{equation}\label{eq:dSMetricConformal}
    			g_{dS} = \Omega^2 \left( -dT^2+dR^2+R^2 \sigma_{A B} d x^{A} d x^{B} \right) \, ,
    		\end{equation}
    		where the conformal factor is given by
    		\begin{equation}
    			\Omega^2 = \frac{4}{(1+\alpha^2(-T^2 +R^2))^2} \, .
    			\label{eq:ConformalFactor}
    		\end{equation}
    		The coordinate transformation between \eqref{eq:dSMetricConformal} and \eqref{eq:deSitterStationary} reads
    		\begin{equation}
    			\begin{cases}
    				T &= \frac{f(r)\sinh{\alpha t}}{\alpha (1+f(r)\cosh{\alpha t})} \, ,\\
    				R &= \frac{ r}{1+f(r)\cosh{\alpha t}} \, ,\\
    				x^A&=x^A \, .
    			\end{cases}
    			\label{eq:CoordTransCMinStatic}
    		\end{equation}  
    		The natural question arises: Is there any ambiguity in choosing a coordinate transformation in which the de Sitter metric is conformally flat? The analysis in appendix \ref{sec:UniqnessDesitter} showed, under reasonable assumptions, that the resulting conditions on coordinate transformation are stringent.

    		In this section, we aim to establish coordinates for de Sitter spacetime for which the description of hopfion-like solution is nearly optimal. In particular, investigating Noetherian charges associated with spacetime symmetries (energy, angular momentum, etc.) is challenging. We will see in section \ref{sec:KillingCharges} that de Sitter spacetime with the metric in the form 
    		\begin{equation}
    			\label{eq:CompactCoordMetric}
    			g=-\frac{1}{\cosh ^2(\alpha \rho)} d t^2+\frac{1}{\cosh ^2(\alpha \rho)} d \rho^2+\alpha^{-2} \tanh ^2(\alpha \rho) d \Omega^2 \, ,
    		\end{equation}
    		allows us to proceed with the energy conveniently. The volume element reads
    		\begin{equation}
    			\sqrt{|\det g|}= \frac{\alpha^{-2} \tanh^2 (\alpha \rho) \sin \theta}{\cosh^2 (\alpha \rho)} \, .
    		\end{equation}
    		\noindent
    		In fact, the transformation between $(t,\rho, x^{A})$ and stationary coordinates (inside cosmological horizon) is one-dimensional
    		\begin{equation}
    			r=\alpha^{-1} \tanh (\alpha \rho) \, .
    		\end{equation}
    		Transformation correlating $(t,\rho, x^{A})$ with coordinates in which the metric is explicitly conformally flat reads
    		\begin{equation}
    			\begin{cases}
    				T &=\frac{ \alpha^{-1} \sinh (\alpha t)}{(\cosh (\alpha t)+\cosh (\alpha \rho))} \, , \\
    				R &=\frac{\alpha^{-1} \sinh (\alpha \rho)}{(\cosh (\alpha t)+\cosh (\alpha \rho))} \, , \\
    				x^{A} &= x^{A} \, .
    			\end{cases}
    		\end{equation}

    		It should be emphasized that in the $(t,\rho, x^{A})$ coordinates, the temporal symmetry vector field has the most straightforward form $\partial_{t}$. In addition to that, the reduced data for the hopfion-like solution, which is a generating function for Maxwell two-form, has a straight form \eqref{eq:PhiHopfiontrho}.  
    		
   	\section{Generalized hopfion-like solutions in de Sitter spacetime\label{sec:GenerHopf}}

   		We obtain a self-dual Maxwell two-form for a hopfion-like solution. The solution is acquired by generalizing reduced data for hopfions in Minkowski space-time.
   		
   		By the self-dual Maxwell two-form we mean a complex two-form $\mathcal{F}_{\mu\nu}$ which is related with a real Maxwell field $F_{\mu\nu}$ as follows
   		\begin{equation}
   			\label{eq:AntiSelfDualDef}
   			\mathcal{F}_{\mu\nu} = F_{\mu\nu} - \imath (\ast F_{\mu\nu})
   			\, ,
   		\end{equation}
   		for details see equation \eqref{eq:SelfDualFDef} in appendix \ref{sec:ReducedDataDeSitter} and the comments nearby. The reduced data is constructed with the help of conformal Yano--Killing (CYK) two-form as
   		\begin{equation}
   			\Phi = \mathcal{F}^{\mu\nu} Q_{\mu\nu} \, .
   		\end{equation}
			For Minkowski space-time and particular choice of $Q=r d t \wedge d r \, ,$ the reduced data is equal to
   			\begin{equation}
   				\Phi=(\mathbf{E} + \imath \mathbf{B}) \cdot \mathbf{R} \, ,
   			\end{equation}
   			where $\mathbf{E}$, $\mathbf{B}$ and $\mathbf{R}$ are electric field, magnetic field and position vector field respectively.
   		
   		In de Sitter space-time,  the construction of self-dual Maxwell two-form from reduced data is described in appendix \ref{sec:ReducedDataDeSitter}.In arbitrary space-time, the essential properties of (CYK) two-form in electromagnetism are given in appendix \ref{sec:ElCYK}.

   		The main point of the section is to find a generalization of hopfions for the de Sitter space-time. The crucial properties needed are the conformal correspondence between Minkowski and de Sitter space-time and conformal covariance of Maxwell theory\footnote{\label{ft:conformal_Maxwell}The conformal group consists of those transformations which leave a metric invariant up to an overall factor, more precisely if we transform coordinates $\tilde{x}^{\mu} \rightarrow x^{\mu}$
   			$$
   			g_{\mu \nu} \mathrm{d} \tilde{x}^\mu \mathrm{d} \tilde{x}^\nu=\Omega(x)^2 g_{\mu \nu} \mathrm{d} x^\mu \mathrm{d} x^\nu .
   			$$
   			Both Maxwell two-form
   			$$
   			F_{\mu \nu}(x) \mathrm{d} x^\mu \mathrm{d} x^\nu=\widehat{F}_{\mu \nu}(\tilde{x}) \mathrm{d} \tilde{x}^\mu \mathrm{d} \tilde{x}^\nu,
   			$$
   			and the standard Maxwell action, 
   			$$
   			S_{A}=- \frac {1}{16 \pi}  \int \mathrm{d}^4 x \sqrt{|-\det g|} g^{\mu \lambda}g^{\nu \sigma} F_{\lambda \sigma} F_{\mu \nu}\, ,
   			$$
   			remain invariant under such transformations, which ensures that Maxwell's equations (Euler--Lagrange equations and $d F=0$) are invariant under conformal transformations.
   		}. The most convenient way to proceed is to analyze the generalization of hopfions in terms of the reduced data. Let us briefly recall the conformal properties of the reduced data. For a general member of a four-dimensional family of Pleba\'nski--Demia\'nski space-times\footnote{In particular, space-times, for which \eqref{diag_Y} holds, are considered.}, the wave equation which describes the evolution of the reduced data \eqref{eq:F_I_P_D} consists of a complex potential and a second-order differential operator
   		\begin{equation}
   			\label{conf_inv_op}
   			\left( \Box-\frac{1}{6} R \right) \Phi \, ,
   		\end{equation}
   		where $R$ is a curvature scalar. The reduced data $\Phi$ rescales conformally in the following way
   		\begin{equation}
   			\label{eq:ConformalCovarianceScalarField}
   			\tilde{\Phi} \to \Omega^{-1} \Phi \, ,
   		\end{equation}
   		under conformal change of a metric ($\widetilde{g}_{\mu\nu}=\Omega^{2} g_{\mu\nu}$). Moreover, the operator \eqref{conf_inv_op} remains invariant under conformal transformation. To be precise, we have 
   		\begin{equation}
   			\left( \widetilde{\square}-\frac{1}{6} \widetilde{R} \right)\widetilde{\Phi}=\Omega^{2}\left( \Box-\frac{1}{6} R \right) \Phi \, , 
   		\end{equation}
   		where tilded objects are calculated with respect to $\Omega^{2} g_{\mu\nu}$. The above conformal properties also apply to our considerations on generalized hopfions in de Sitter space-time; see the conformal factor \eqref{eq:ConformalFactor} and the comments nearby.
   		
   		As a next step, we recall the reduced data for hopfions\footnote{The equation \eqref{eq:Hopfionreddata} is given as (2.12) in \cite[page 5]{Smolka_2018}. The classical hopfion solution, obtained by Ra\~nada, is the case with $l=1$. } --- one-parameter, denoted by $l$, family of electromagnetic solutions in Minkowski space-time (see \cite[page 5]{Smolka_2018}). In spherical coordinates\footnote{We use the exact representation of two-dimensional spherical metric \eqref{eq:ExactTwoForm}.}, the hopfions read
   		\begin{equation}
   			\Phi_l = \frac{r^l \sin^l \theta(\cos \varphi+\imath \sin \varphi)^l}{\left[ -(t-\imath)^2 + r^2 \right]^{l+1}} \, .
   			\label{eq:HopfionsMinkowski}
   		\end{equation}
   		The scalar field fulfills the wave equation
   		\begin{equation}
   			\Box \Phi_l=0 \, .
   			\label{eq:waveMinkowski}
   		\end{equation}
   		For a flat space-time, the curvature scalar vanishes, so the operator in the equation \eqref{eq:waveMinkowski} is in the form of \eqref{conf_inv_op}. By rescaling the coordinates $(t,r,\theta,\varphi)\to (\frac{t}{\widehat{\beta}},\frac{r}{\widehat{\beta}},\theta,\varphi)$, we receive a scalar parameter freedom in the solution. Because of the symmetries of Minkowski space-time, the rescaling is completely redundant. However, it matters when we pass into de Sitter space-time. The rescaled solution reads
   		\begin{equation}
   			\label{eq:Hopfionreddata}
   			\Phi_l = \frac{ \widehat{\beta}^{l+2} r^l \sin^l \theta(\cos \varphi+\imath \sin \varphi)^l}{\left[ -(t-\imath \widehat{\beta})^2 + r^2 \right]^{l+1}}
   			\, .
   		\end{equation}

   		According to the conformal covariance of electromagnetism, the rescaled reduced data for de Sitter space-time in explicitly conformally flat $(T,R,\theta,\varphi)$ coordinates \eqref{eq:CoordTransCMinStatic} reads
   		\begin{equation}
   			\Phi_l=	
   			\frac{ R^l \sin^l \theta(\cos \varphi+\imath \sin \varphi)^l \left(\alpha^2 R^2-\alpha^2 T^2+1\right)}{\left[ -(T-\imath \widehat{\beta})^2 + R^2 \right]^{l+1}}	
   			\, .
   		\end{equation}  
   		At first glance, one can think that the above transformed reduced data is wholly defined. However, some subtleties relate to ambiguities of conformal flatness of the de Sitter space-times (see section \ref{sec:UniqnessDesitter}). We continue analyzing reduced data in flat space limit $\alpha \to 0$.

   		\subsection{Flat space limit\label{sec:FlatSpaceLimit}}
   		We wish to choose the constant $\widehat{\beta}$, such that in the limit $\alpha\to0$, the constructed reduced data overlaps with the hopfion scalar in the Minkowski space. Superposing coordinate transformation \eqref{eq:CoordTransCMinStatic} with conformal covariance \eqref{eq:ConformalCovarianceScalarField}, we obtain a formula for reduced data in static coordinates
   		\begin{equation}
   			\Phi_l = \frac{\widehat{\beta}^{l+2}(\alpha r)^l \sin^l \theta(\cos \varphi+\imath \sin \varphi)^l}
   			{\left[ 
   				(1+\widehat{\beta}^2) + (1-\widehat{\beta}^2)\sqrt{1-\alpha^2r^2}\cosh{\alpha t} + 2\imath\widehat{\beta}\sqrt{1-\alpha^2r^2}\sinh{\alpha t}
   				\right]^{l+1}}
   			\, .
   			\label{eq:ReducedDataStaticCoords}
   		\end{equation}
   		Let us observe that $\widehat{\beta}$ should depend on $\alpha$; otherwise, the denominator in \eqref{eq:partial_reduced_data} is constant in the limit. Moreover, rearranging the denominator in $\Phi_l$, up to a constant, the scalar field reads 
   		\begin{equation}
   			\label{eq:partial_reduced_data}
   			\Phi_l = \frac{r^l \sin^l \theta(\cos \varphi+\imath \sin \varphi)^l}
   			{\left[ 
   				(1+\alpha^2 \widehat{\beta}^2) + (1-\alpha^2 \widehat{\beta}^2)f(r)\cosh{\alpha t} + 2\imath \alpha \widehat{\beta} f(r)\sinh{\alpha t}
   				\right]^{l+1}} \, .
   		\end{equation}
   		It is clear, that $\widehat{\beta}$ should be of order $\alpha^{-2}$. Calculating the limit $\alpha \to 0$, we find
   		\begin{equation}
   			\label{eq:beta_param}
   			\widehat{\beta}=\frac{1}{\alpha^2 \beta} \, ,
   		\end{equation} 
   		then, up to overall constant factor, the reduced data reads
   		\begin{equation}
   			\Phi_l =\frac{r^l  \sin^l \theta(\cos \varphi+\imath \sin \varphi)^l}
   			{\left[ \Big( (\beta^2 \alpha^2 - 1) \cosh(\alpha t)+2 \imath  \beta \alpha \sinh (\alpha t)\Big) f(r)+\alpha^2 \beta^2+1\right]^{l+1}}
   			\, .
   			\label{eq:reduced_data_tb}
   		\end{equation}
   		Passing to the limit $\alpha \to 0$, we obtain, up to the overall constant factor, the equation \eqref{eq:Hopfionreddata} with $\beta=\widehat{\beta}$.

   		In $(t, \rho, \theta, \varphi)$ coordinates\footnote{We use coordinates in which the metric has the form \eqref{eq:CompactCoordMetric} with explicit representation of angular coordinates \eqref{eq:ExactTwoForm}.}, the reduced data for generalized hopfions\footnote{Obtained up to an overall constant factor.} has a compact form
   		\begin{equation}
   			\Phi_{l}=\frac{ \sinh (\alpha \rho)^{l} \cosh (\alpha \rho)  \sin^l \theta e^{\imath l \varphi}}{[(1+\beta^2 \alpha^2)(\cosh (\alpha \rho)-\cosh (\alpha t-\imath \upsilon))]^{l+1}} \, ,
   			\label{eq:PhiHopfiontrho}
   		\end{equation}
   		where $\upsilon \in ]0,2 \pi[$ is defined by
   		\begin{equation}
   			\begin{cases}
   				\sin \upsilon &= \frac{2 \beta \alpha}{1+\beta^2 \alpha^2} \, ,\\
   				\cos \upsilon &= \frac{1-\beta^2 \alpha^2}{1+\beta^2 \alpha^2} \, .
   			\end{cases}
   			\label{eq:upsilonPhase}
   		\end{equation}

   		A careful reader may challenge the usage of reduced data in our analysis and ask if a direct transformation of Maxwell's two-form is not the simpler way to achieve generalization of hopfions to de Sitter space-time. In our context, the reduced data has a few powerful assets. First, the reduced data for generalized hopfion-like solutions is trivial from the point of view of spherical foliation. A detailed explanation is the following. The de Sitter space-time is spherically symmetric. A natural foliation of spheres can be introduced. All used coordinate systems have implemented such foliation, spanned by $(\theta,\varphi)$ coordinates. The reduced data for a generalized hopfion-like solution is proportional to l-pole (one multi-pole)\footnote{By l-pole, we mean only one term $Y_{l}$ in multi-pole expansion. }, denoted by $Y_{l}$
   		\begin{equation}
   			\Phi_l \propto Y_{l} \, .
   		\end{equation}   
   		The essential step in the recovery procedure for hopfions (appendix \ref{sec:ReducedDataDeSitter}) is to invert the two-dimensional Laplace operator $\mathbf{\Delta}$. $Y_{l}$ is an eigen-function of $\mathbf{\Delta}$, so $\mathbf{\Delta}^{-1} Y_{l}$ is equivalent to division by $- l (l+1)$. Moreover, $\mathbf{\Delta}^{-1} Y_{l}$ occurs in most physical functionals (like energy or helicity). Two of us have analyzed such functionals in Minkowski space-time. For example, see sections 2.3, 3.2, and 3.3 in \cite{Smolka_2018}. Analogically, for Maxwell two-form, such functionals can be rewritten regarding reduced data (appendix \ref{sec:ReducedDataDeSitter}) with the help of identities from appendix \ref{sec:spher_ident}. In other words, we can reduce the analysis for the generalized hopfion-like solutions to a two-dimensional picture (normal to the spherical foliation) in this way. Moreover, we highlight that the procedure presented in sections \ref{sec:GenerHopf} and \ref{sec:FlatSpaceLimit} contains in addition to coordinate change and conformal transformation of the field\footnote{According to \eqref{eq:ConformalCovarianceScalarField}.}, also an inversion-like transformation of the parameter $\widehat{\beta}$, see \eqref{eq:beta_param}. So, there are subtleties that are required to be taken into account here.
   		The recovery procedure, described in appendix \ref{sec:ReducedDataDeSitter}, enables one to receive self-dual Maxwell two-form for the generalized hopfion-like solutions from $\Phi_l$. We restrict ourselves only to the case $l=1$ which is a generalization of Ra\~nada classical hopfion from Minkowski space-time. The self-dual Maxwell two-form in stationary coordinates reads
   		\begin{eqnarray}
   			\mathcal{F}_{H}&=&e^{\imath \varphi} \left\{ 2 \chitrho \sin \theta \left(\frac{1}{r^2} d t \wedge d r+\imath  \sin \theta  d \theta \wedge d \varphi \right) \right.
   			\nonumber \\
   			& & -\left(\partial_{t}\chitrho-f(r)^2\partial_{r} \chitrho \cos \theta   \right)\left(d t \wedge d \theta -\frac{\imath  \sin \theta }{f(r)^{2}} d r \wedge d \varphi \right)
   			\label{eq:FHtr}
   			\\
   			& & \left. - \left(f(r)^2\partial_{r} \chitrho - \partial_{t} \chitrho \cos \theta \right)\left(
   			\frac{1}{f(r)^2}d r \wedge d \theta-\imath \sin \theta d t \wedge d \varphi\right)  \right\}
   			\, ,
   			\nonumber
   		\end{eqnarray}
   		where $\chitrho$ is defined by the relation
   		\begin{equation}
   			r \Phi_{1}=\chitrho \sin \theta e^{\imath \varphi} \, .
   		\end{equation}
   		Explicitly, we have
   		\begin{equation}
   			\chitrho(t,r)=\frac{r^2}{(1+\beta^2 \alpha^2)^2(1-f(r)\cosh (\alpha t-\imath \upsilon) )^2} \, ,
   			\label{eq:chitrho}
   		\end{equation}
   		together with $\upsilon$ defined by \eqref{eq:upsilonPhase}. Equivalently, $\mathcal{F}_{H}$ in $(t,\rho,\theta,\varphi)$ is given by
   		\begin{eqnarray}
   			\mathcal{F}_{H}&=&e^{\imath \varphi} \left\{ 2 \chitrho \sin \theta \left(\frac{\alpha^2}{\sinh^2(\alpha \rho)} d t \wedge d \rho+\imath  \sin \theta  d \theta \wedge d \varphi \right) \right.
   			\nonumber \\
   			\label{eq:hopftwoformrho}
   			& & -\left(\partial_{t}\chitrho-\partial_{\rho} \chitrho \cos \theta   \right)(d t \wedge d \theta -\imath  \sin \theta d \rho \wedge d \varphi )  \\
   			& & \left. - \left(\partial_{\rho} \chitrho - \partial_{t} \chitrho \cos \theta \right)(d \rho \wedge d \theta-\imath \sin \theta d t \wedge d \varphi)  \right\}
   			\, ,
   			\nonumber
   		\end{eqnarray}
   		where, analogically, $\chitrho$ is given by \eqref{eq:chitrho}. We find the explicit form
   		\begin{equation}
   			\chitrho(t,\rho)=\frac{ \sinh^2(\alpha \rho)}{(1+\beta^2 \alpha^2)^2(\cosh (\alpha \rho)-\cosh (\alpha t-\imath \upsilon))^{2}}
   			\, ,
   			\label{eq:chitrho2}
   		\end{equation}
   		with $\upsilon$ defined by \eqref{eq:upsilonPhase}.
	
   	\section{Topological properties of the solutions\label{sec:TopoProperties}}

   	 	\begin{wrapfigure}{R}{0.5\textwidth}
   	 		\begin{center}
   	 			\includegraphics[width=0.48\textwidth]{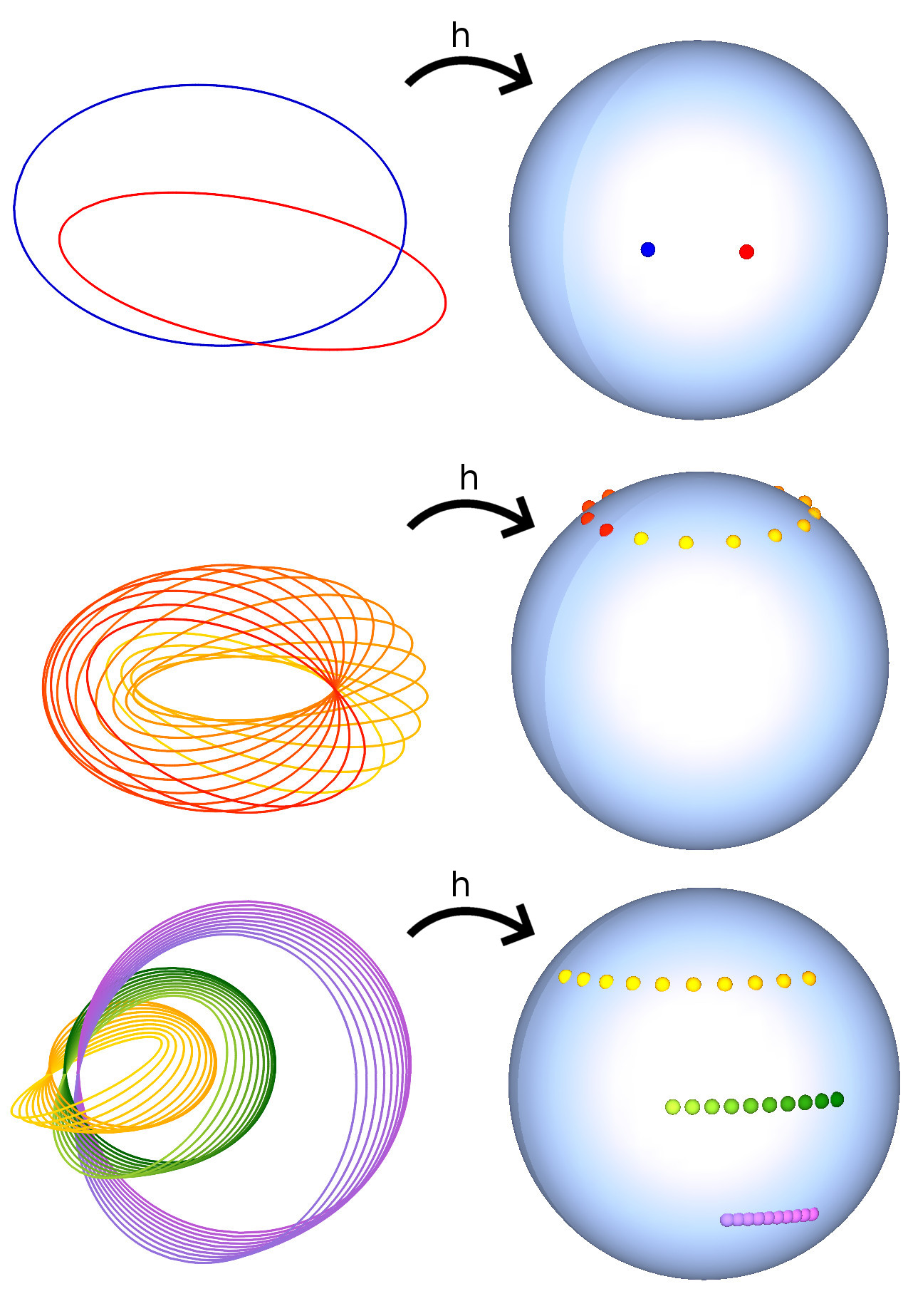}
   	 		\end{center}
   	 		\caption{ \small The structure of Hopf fibration of $\mathbb{S}^3$. A torus can be constructed out of circles (fibers) so that no two circles cross and each circle is linked to every other one. In the left picture in the second row, each circle in such a configuration wraps once around each circumference of the torus.
   	 			The last picture in the left column contains parts of nesting such tori into one another, the whole of three-dimensional space, including the point at $R=\infty$ $\left(\mathbb{R}^{3} \cup \infty \sim \mathbb{S}^{3}\right)$, can be filled with linked circles. The right column contains corresponding images of fibers from the left column under the Hopf projection.  
   	 		}
   	 		\label{fig:Hopf_fibration}
   	 	\end{wrapfigure}

   	 	We begin our topological analysis by briefly reviewing the classical properties of hopfions in Minkowski space. Subsequently, we determine whether the results can be extended on de Sitter spacetime as conformally flat spacetime.
   	 	
   	 	Following \cite{irvine2008linked,ranada_1990knot,Smolka_2018}, hopfion is a 'solitonary' solution of Maxwell theory\footnote{In general, the hopfion solutions for linearized gravity or even spin-N field are known.} which has a rich topological structure related to the Hopf fibration. The Hopf fibration is the simplest non-trivial fibration of a three-dimensional sphere. The projection of the Hopf fiber bundle, called the Hopf map $h$, is a surjective map sending great circles on $\mathbb{S}^{3}$ (fibers) to points on $\mathbb{S}^2$, see Figure \ref{fig:Hopf_fibration}. These circles weave nested toroidal surfaces, and each is linked with every other circle exactly once, creating the characteristic structure of Hopf fibration. This structure can be easily found in the integration curves of the hopfion solution. The closed, linked field lines of hopfions propagate without intersections along the light cone. 
   	 	
   	 	The conformal flatness of de Sitter spacetime enables one to generalize some results obtained for Minkowski spacetime to de Sitter spacetime. The conformal methods have, nevertheless, some noticeable obstructions. First, the generalized structure needs to pose conformal covariance, which is understood here as a clear, relatively simple conformal transformation law. In our case, Maxwell's theory is preserved under conformal transformation (recall footnote \ref{ft:conformal_Maxwell} on page \pageref{ft:conformal_Maxwell}). Second, conformal flatness does not ensure the compatibility of global structures of spacetime. For example, the topological structures of de Sitter spacetime and Minkowski spacetime differ. In particular, the cosmological horizon is suspected of presenting new behavior when the hopfion wave approaches them. The research below aims to analyze the interference of topological structures between the hopfion solution and de Sitter spacetime.

   	 	We wish to analyze the topological properties of the generalized hopfion-like solutions $(l=1)$. In particular, integral curves of electromagnetic fields on
   	 	\begin{equation}
   	 		\Sigma_{0}:=\{(t,\rho,\theta, \varphi): t=0\} \, ,
   	 	\end{equation}
   	 	are investigated for the interior neighborhood of the horizon. For convenience, the electric $\textbf{E}$ and magnetic $\textbf{B}$ vector fields are combined in a complex-valued vector field
   	 	\begin{equation}
   	 		\textbf{Z}=\textbf{E}+\imath \textbf{B} \, ,
   	 	\end{equation}
   	 	known as Riemann--Silberstein vector. The components of Riemann--Silberstein field $\textbf{Z}=Z^{i} \partial_{i}$ are defined by Maxwell field as  
   	 	\begin{equation}
   	 		Z^i = \frac{\mathcal{F}^{t i}}{\sqrt{-g^{t t}}} \, .
   	 	\end{equation}
   	 	Using \eqref{eq:FHtr}, we find the explicit form of the Riemann--Silberstein vector field for the generalized hopfion-like solutions $(l=1)$ on $\Sigma_{0}$
   	 	\begin{equation}
   	 		\label{eq:field_radial}
   	 		\begin{aligned}
   	 			\lefteqn{ \textbf{Z}=\frac{f(r) e^{\imath \varphi}}{r^2} \bigg\{ -2 \chitrho \sin \theta \,   \partial_{r} } & \\
   	 			& +\frac{\left(\partial_{t}\chitrho-f(r)^2\partial_{r} \chitrho \cos \theta   \right)}{ f(r)^2} \, \partial_{\theta} 
   	 			\phantom{xxx} 	
   	 			\\
   	 			& - \frac{\imath \left(f(r)^2\partial_{r} \chitrho - \partial_{t} \chitrho \cos \theta \right)
   	 			} {f(r)^2 \sin \theta } \, \partial_{\varphi}   \bigg\}
   	 			\, , 
   	 		\end{aligned}
   	 	\end{equation}
   	 	where $\chitrho$ and its derivatives are obtained on $\Sigma_{0}$
   	 	\begin{equation*}
   	 		\begin{aligned}
   	 			\chitrho|_{t=0}&=\frac{r^2}{\left[f(r) (\alpha^2 \beta^2-1)+\alpha^2 \beta^2+1 \right]^2} \, ,\\
   	 			\partial_{t} \chitrho|_{t=0}&=\frac{-4 \imath \alpha^2 r^2 f(r) \beta}{\left[f(r) (\alpha^2 \beta^2-1)+\alpha^2 \beta^2+1 \right]^3} \, , \\
   	 			\partial_{r} \chitrho|_{t=0}&=\frac{2 r\left[f(r) (\alpha^2 \beta^2+1)+\alpha^2 \beta^2-1 \right]}{f(r)\left[f(r) (\alpha^2 \beta^2-1)+\alpha^2 \beta^2+1 \right]^3} \, . \\
   	 		\end{aligned}
   	 	\end{equation*}

   	 	To investigate the field lines analytically, we introduce on $\Sigma_{0}$ radial rescaling, given by
   	 	\begin{equation}
   	 		r = \frac{4 R}{(4+\alpha^2 R^2)} 
   	 		\, ,
   	 	\end{equation}
   	 	where $R=\sqrt{x^2+y^2+z^2}$. Together with the Cartesian-like transformation
   	 	\begin{equation}
   	 		\begin{cases}
   	 			x=R \sin \theta \cos \varphi \, ,\\
   	 			y=R \sin \theta \sin \varphi\, ,\\
   	 			z=R \cos \theta \, ,
   	 		\end{cases}
   	 	\end{equation}
   	 	the induced three-dimensional metric is expressed as
   	 	\begin{equation}
   	 		g|_{\Sigma_{0}} = \frac{16}{(4+\alpha^2 R^2)^2}\left(d x^2+d y^2+ d z^2\right)
   	 		\, .
   	 	\end{equation}
   	 	The associated Cartesian-like set of coordinates is related with 
   	 	toroidal ones by
   	 	\begin{equation}
   	 		\begin{split}
   	 			z &= \beta\frac{\sinh{\eta}}{\cosh{\eta}-\cos{\sigma}}\cos{\phi} \, ,\\
   	 			x &= \beta\frac{\sinh{\eta}}{\cosh{\eta}-\cos{\sigma}}\sin{\phi} \, ,\\
   	 			y &= \beta\frac{\sin{\sigma}}{\cosh{\eta}-\cos{\sigma}} 
   	 			\, ,\\
   	 		\end{split}
   	 	\end{equation}
   	 	leads to a simple form of the electric field $E^i \partial_{i}$ on $\Sigma_{0}$:
   	 	\begin{equation}\label{eq:field_toroidal}
   	 		\begin{split}
   	 			E^\eta &= 0
   	 			\, ,\\
   	 			E^\sigma &= -\frac{ (\cosh{\eta}-\cos{\sigma}) \left(\left(1+\alpha^{2} \beta^2 \right) \cosh{\eta} - 
   	 				\left(1-\alpha^{2} \beta^2 \right) \cos{\sigma}\right)^2}
   	 			{32 \beta \cosh^3{\eta}}
   	 			\, ,\\
   	 			E^\phi &= \frac{ (\cosh{\eta}-\cos{\sigma}) \left(\left(1+\alpha^{2} \beta^2 \right) \cosh{\eta} - 
   	 				\left(1-\alpha^{2} \beta^2 \right) \cos{\sigma}\right)^2}
   	 			{32 \beta \cosh^3{\eta}}
   	 			\, .
   	 		\end{split}
   	 	\end{equation}

   	 	The $\eta$-component of the electric field is zero, so the field lines are confined to two-dimensional surfaces parameterized by $\eta$, tori given by the family of equations in Cartesian-like coordinates
   	 	\begin{equation}\label{eq:tori}
   	 		y^2 + \left( \sqrt{x^2+z^2} - \beta\coth{\eta} \right)^2 =
   	 		\frac{\beta^2}{\sinh^2{\eta}}
   	 		\, .
   	 	\end{equation}
   	 	The magnetic field manifests a similar toroidal structure but rotated in the $XY$ plane by 90 degrees. 
   	 	Generalized hopfions in de Sitter spacetime share the remarkable topological properties of classical hopfions in Minkowski spacetime (see \cite{Thompson_2015}), in that the electric and magnetic field lines on $\Sigma_{0}$ form toroidal structures which correspond to Hopf fibration. 
   	 	\begin{figure}[]
   	 		\centering
   	 		\includegraphics[width=0.48\textwidth]{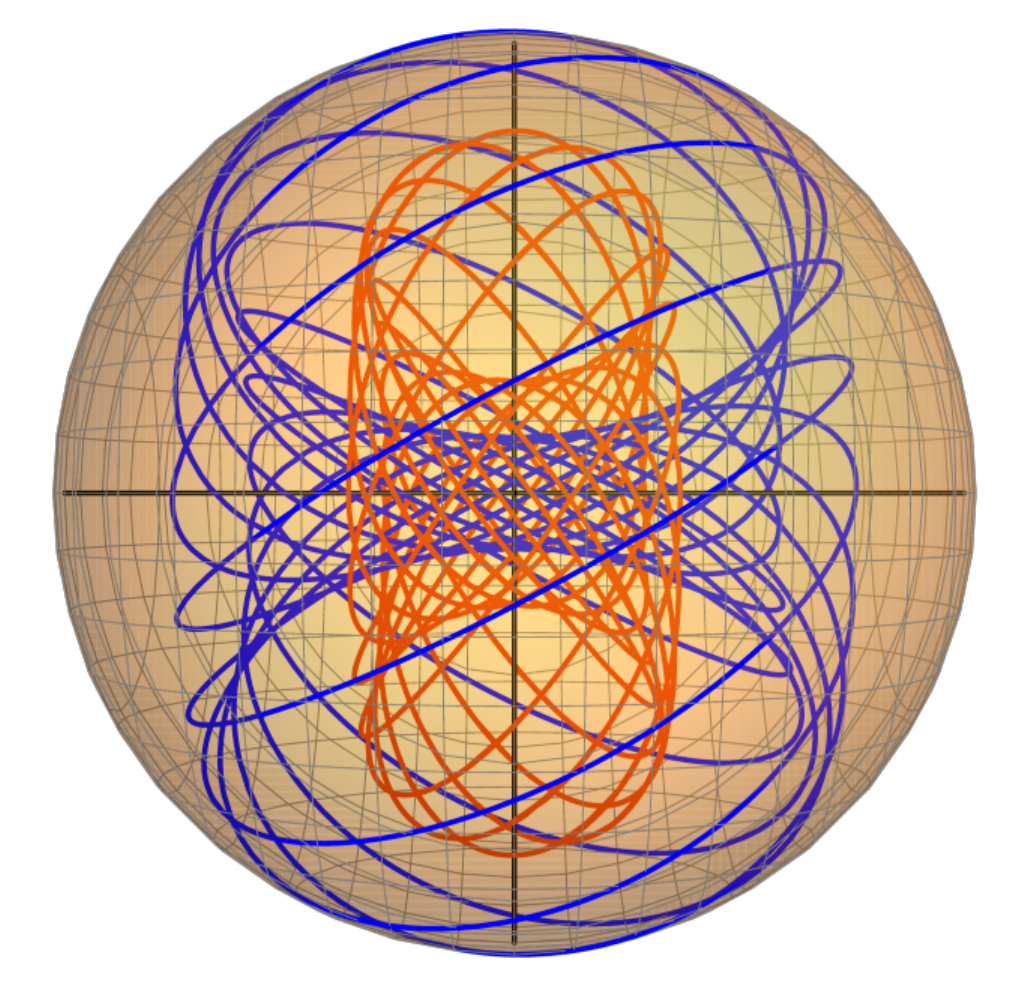}
   	 		\includegraphics[width=0.48\textwidth]{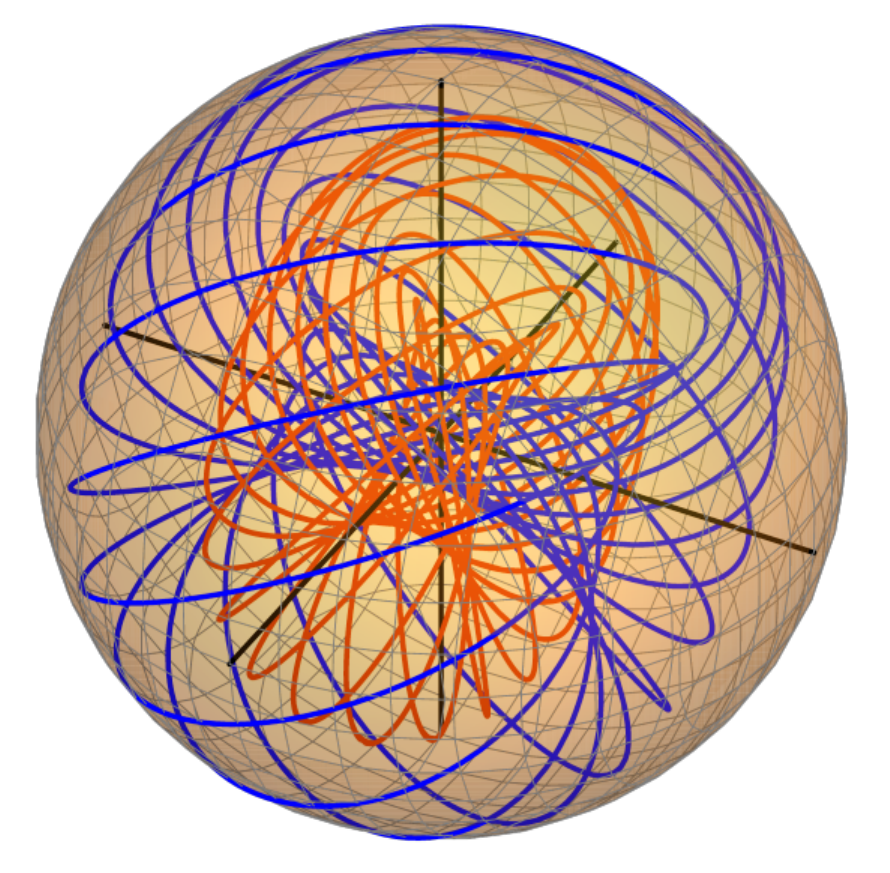}
   	 		\caption{Field lines for the electric field, forming toroidal structures inside the cosmological horizon. Both figures present a particular choice of integral curves in static coordinates from different perspectives. The sphere represents the cosmological horizon.}
   	 		\label{fig:inside_lines}
   	 	\end{figure}
   	 	
   	 	The structure of field lines, presented in the figure \ref{fig:inside_lines}, forms a system of linked, closed circles. Each two lines are linked only once. When choosing the initial data (starting point) on the toroidal surface, the integral curves do not leave the torus. The field line on the symmetry axis is analytically investigated in section \ref{sec:symmetryline}.

   	 	\subsection{Field line which reaches the horizon in finite time \label{sec:symmetryline}}
   	 	The structure of the electric field given by formulas \eqref{eq:field_radial} allows us to investigate two distinguished field lines on $\Sigma_{0}$, corresponding to two degenerate tori---the OX axis, and a ring in the YZ plane. The electric field at $t=0$ in static coordinates is explicitly given by:
   	 	\begin{equation}
   	 		\begin{split}
   	 			E^r &=\frac{4  \alpha^4 \beta^4  \sqrt{1-\alpha^2 r^2} \cos \theta}{\left(\alpha^{2} \beta^2 +1+\left(\alpha^{2} \beta^2 -1\right) \sqrt{1-\alpha^2 r^2}\right)^2} \, ,\\
   	 			E^{\theta} &=-\frac{4 \alpha^4 \beta^4 \left(\left(-1+\alpha^{2} \beta^2 \right)+\left(1+\alpha^{2} \beta^2 \right) \sqrt{1-\alpha^2 r^2}\right)}{r\left(1+\alpha^{2} \beta^2 +\left(-1+\alpha^{2} \beta^2 \right) \sqrt{1-\alpha^2 r^2}\right)^3} \sin \theta \, ,\\
   	 			E^{\phi} &=\frac{8 \alpha^6 \beta^5 }{\left(1+\beta^2 \alpha^2+\left(-1+\beta^2 \alpha^2\right) \sqrt{1-\alpha^2 r^2}\right)^3} \, .
   	 		\end{split}
   	 	\end{equation}
   	 	Thus, setting $\theta=0$, finding the integral curve reduces to a one-dimensional problem. The equation 
   	 	\begin{equation}
   	 		\dot{r}(s) = E^r(r(s))
   	 	\end{equation}
   	 	is readily solved by the separation of variables. A direct calculation shows that a field line starting at the origin reaches the horizon in finite time, precisely given by:
   	 	\begin{equation}
   	 		s_{horizon} = \frac{(8+3 \pi ) \alpha^4 \beta^4 +2 \pi  \alpha^2 \beta^2+ (3 \pi -8)}{16 \alpha^5 \beta^4}
   	 		\, .
   	 	\end{equation}
   	 	
   	 	We can set $\theta=\frac{\pi}{2}$. The radial component vanishes, so as long as the $\theta$-component of the field remains zero, a circular integral line in the XY-plane would arise. Setting $E^\theta=0$, we obtain an equation for the radius of said field line
   	 	\begin{equation}
   	 		r_{focus} = \frac{2 \beta}{1+\alpha^2 \beta^2} \, .
   	 	\end{equation}
   	 	This field line in conformal coordinates coincides with the focal ring of the tori given by \eqref{eq:tori}.
   	 	
   	\section{Noether charges}
   		\subsection{Charges related with symmetries of the spacetime\label{sec:KillingCharges}}
			
			In Minkowski space-time, the hopfion solution is characterized by classical conserved quantities associated with space-time symmetries. In particular, the energy functional is given by 
			\begin{equation}
				\mathit{E}= \frac{1}{2} \int(\mathbf{E}^2+\mathbf{B}^2) d V \, ,
				\label{eq:EnergyMinkowski}
			\end{equation}
			where $\mathbf{E}$ and $\mathbf{B}$ are the electric and magnetic fields, respectively. The total angular momentum is defined as
			\begin{equation}
				\mathbf{J}= \int \mathbf{r} \times(\mathbf{E} \times \mathbf{B}) d V \, ,
				\label{eq:AngMomMinkowski}
			\end{equation}
			where $\mathbf{r}$ is a position vector. By $\times$ we denote a standard vector product in flat space-time. The integrands in \eqref{eq:EnergyMinkowski} and \eqref{eq:AngMomMinkowski} are energy and angular momentum densities, respectively.
			
			For de Sitter space-time, the article \cite{2023PTCTSHamiltonianCharges} delves into the study of conserved quantities associated with Noether symmetries. Additionally, in the section \ref{sec:TopoCharges}, we wish to analyze the topological charges related to the duality symmetry. For this purpose, a brief revision of the hamiltonian framework on the de Sitter background is necessary. This unified treatment allows for examining all conserved objects discussed in the paper.
			
			In our signature, the lagrangian for source-free, real electromagnetic fields in vacuum reads 
			\begin{equation}\label{23V21.1}
				\mcL (\partial A_{\mu}) =- \frac {1}{16 \pi}  \sqrt{|\det g|}
				\big( 
				g^{\mu\nu} g^{\alpha\beta} F_{\mu \alpha} F_{\nu\beta}
				\big)
				\,.
			\end{equation}
			The Euler--Lagrange equation of field evolution resulting from the action 
			\begin{equation} 
				S = \int d^4 r \,
				\mcL (\partial A_{\mu})
			\end{equation} 
			is
			\begin{equation} 
				\partial_\beta\,F^{\alpha\beta}=0 \, ,
				\label{eq:MaxwellEq1}
			\end{equation}
			together with the second exterior derivative of potential $d d A=0$, which implies
			\begin{equation}
				\partial_\beta (\ast F)^{\alpha\beta}=0 \, ,
				\label{eq:MaxwellEq2}
			\end{equation}
			yields a form of the Maxwell equations.
			Denoting $\partial_\nu A_\mu$ by $A_{\mu,\nu}$, the canonical momentum density reads
			\begin{eqnarray}
				\pi^{\alpha \beta}=\frac{\partial\mcL}{\partial A_{\beta,\alpha} }=-\frac{1}{4 \pi} \sqrt{|\det g|} F^{\alpha \beta} \, .
				\label{29V21.t2}
			\end{eqnarray}
			The standard hamiltonian current, which we will denote by $\mcH^\mu_c$,   is  defined as
			\begin{eqnarray}
				\mcH^\mu_c[X]
				&:= & \frac{\partial \mcL}{\partial A_{\beta,\mu} } \Lie_{X} A_{\beta}
				-
				\mcL
				X^\mu
				\nonumber
				\\
				& = &
				- \frac {1}{4 \pi}  \sqrt{|\det g|}
				\big(   F^{\mu \beta}\Lie_{X} A_{\beta}
				- \frac 1 4 F^{\alpha \beta} F_{\alpha\beta} X^\mu
				\big)
				\, ,\label{23V21.t1old}
			\end{eqnarray}
			where $\Lie_X A$ denotes Lie derivative of a covector field $A$. The problem with the hamiltonian \eqref{23V21.t1old} is its gauge dependence, which can be fixed by
			replacing $\Lie_XA$ by\footnote{$\myLie_{X}A_{\mu}$ is a natural definition for the Lie derivative of a connection one form on a $U(1)$ principal bundle.}
			\begin{equation}
				\label{24V21.t4}
				\myLie_{X}A_{\mu} := X^{\nu}F_{ \nu \mu} \, ,
			\end{equation}
			and defining
			\begin{eqnarray}
				\mcH^\mu [X]
				& := &   \frac{\partial \mcL}{\partial A_{\beta,\mu} } \myLie_{X} A_{\beta}
				-
				\mcL
				X^\mu
				\nonumber
				\\
				&= &      -\frac{1}{4 \pi} \sqrt{|\det g|}
				\Big(F^{\mu \beta} \myLie_{X} A_{\beta}
				-
				\frac {1}{4}
				\big(
				F^{\nu \beta} F_{\nu\beta}
				\big)
				X^\mu
				\Big)
				\nonumber
				\\
				&=&
				-\frac{1}{4 \pi} \sqrt{|\det g|}
				\Big(F^{\mu \beta} X^{\alpha} F_{\alpha \beta}
				-
				\frac {1}{4}
				\big(
				F^{\nu \beta} F_{\nu\beta}
				\big)
				X^\mu
				\Big)
				\,.
				\phantom{xxxx}
				\label{23V21.t1}
			\end{eqnarray}
			In \cite{2023PTCTSHamiltonianCharges}, one of us obtained our final formulae for the Lie derivative of the hamiltonian of a sourceless electromagnetic field. Assuming $X$ and $Y$ are Killing vector fields, we have:
			\begin{equation}
				\label{eq:FluxDef}
				\Lie_{Y} \mcH^{\mu} [X]
				=
				-\frac{1}{2 \pi} \sqrt{|\det g|} \nabla_{\sigma} \left[Y^{[\sigma} F^{\mu] \alpha}
				X^{\kappa}F_{\kappa \alpha} - \frac{1}{4} Y^{[\sigma} X^{\mu]} F^{\alpha \beta} F_{\alpha \beta} \right] \, .
			\end{equation}

			We wish to determine the hamiltonian (Noether) charges associated with space-time's temporal and rotational symmetry. Both symmetries are the Killing vector fields, defined in section \ref{sec:KillingVF}. To continue the analysis, we observe that the lagrangian density \eqref{23V21.1} for hopfion-like solution vanishes at any point
			\begin{equation}
				\mcLHop=0 \, .
			\end{equation}
			The hamiltonian density \eqref{23V21.t1old} is linear in the hamiltonian vector field $X$, restricting ourselves to Killing vector fields; energy and angular momentum currents yield
			\begin{eqnarray}
				\mcH^{\mu} [\KilT]&=&-\frac{1}{4 \pi} \sqrt{|\det g|}
				F^{\mu \beta} F_{\alpha \beta} \left(\partial_{t}\right)^{\alpha}
				\,,
				\label{eq:Ecurrent}
				\\
				\mcH^{\mu} [\KilR]&=&-\frac{1}{4 \pi} \sqrt{|\det g|}
				F^{\mu \beta} F_{A \beta}  \varepsilon^{A B} \zspaceD_{B}(R_i \wtx ^i)
				\,,
				\label{eq:Jcurrent}
			\end{eqnarray}
			where $\varepsilon^{A B}$ is a two-dimensional Levi--Civita tensor\footnote{ Two-dimensional Levi--Civita tensor is defined as
				\begin{equation}
					\sqrt{\det \sigma_{A B}}\varepsilon^{3 4}=1 \, ,
				\end{equation}
				in spherical coordinates $(\theta, \phi)$ we obtain $\varepsilon^{\theta \phi}=\frac{1}{\sin \theta} \, .$
			}
			and the fields
			\begin{equation}
				\wtx ^{i}:=\frac{x^{i}}{r}
				\label{eq:nfields_def}
			\end{equation}
			are scalar fields with additional indices. Thus, $\wtx^i$ is viewed as a scalar on $\mathbb{S}^2$ for covariant derivation. $\wtx^i$ form a basis of the space of dipole function ($\ell=1$ spherical harmonics). $R_{i}$ is a set of constant coefficients. For an explicitly given representation of dipole functions
			\begin{align}
				\wtx^{x}=& \sin \theta \cos \varphi \, ,\\
				\wtx^{y}=& \sin \theta \sin \varphi \, , \\
				\wtx^{z}=& \cos \theta \, , 
			\end{align}
			and the unit sphere metric \eqref{eq:ExactTwoForm}, the associated basis of vector fields reads
			\begin{align}
				\label{eq:KilRx}
				\KilRx=\varepsilon^{A B} \zspaceD_{B}(\wtx^x) \partial_{A}&=   -\sin \varphi \partial_{\theta} -\cot \theta \cos \varphi \partial_{\varphi} \, , \\
				\KilRy=\varepsilon^{A B} \zspaceD_{B}(\wtx^y) \partial_{A}&=   \cos \partial_{\theta} -\cot \theta \sin \varphi \partial_{\varphi} \, , \\
				\KilRz=\varepsilon^{A B} \zspaceD_{B}(\wtx^z) \partial_{A}&=   \partial_{\varphi} \, .
				\label{eq:KilRz}
			\end{align}
			Let us recall that the time components of vector field densities \eqref{eq:Ecurrent} and \eqref{eq:Jcurrent} are generalized counterparts of energy density \eqref{eq:EnergyMinkowski} and angular momentum density \eqref{eq:AngMomMinkowski}, respectively. 
			
			Before we pass to the analysis of charge integrals, some comments about integration might be in order. By $\mathcal{B}_{R}$ we denote a three-dimensional ball of radius $R$
			\begin{equation*}
				\mathcal{B}_{R}=\{(r,\theta,\varphi):r\leq R\} \, .
			\end{equation*} 
			In particular, we introduce 
			\begin{equation*}
				\mathcal{B}_{\Hor}=\left\{(r,\theta,\varphi):r\leq \frac{1}{\alpha}\right\} \, .
			\end{equation*} 
			In addition to this, we use the following indications for measure (volume form):
			\begin{eqnarray}
				&
				dS_\mu := \partial_\mu \rfloor dx^{ 0}\wedge \cdots \wedge dx^n\,,
				\quad
				dS_{\mu\nu} := \partial_\mu \wedge \partial_\nu \rfloor dx^{ 0}\wedge \cdots \wedge dx^n
				\equiv -\partial_\mu \rfloor dS_\nu
				\,,
				\phantom{xxx}
				&
				\nonumber%
				\\
				&
				d \Sigma_{t} = \sqrt{\det g|_{\Sigma_{t}}} \; d \rho \wedge d \theta \wedge d \varphi
				\,, \qquad
				d\sigma = \sqrt{\det \sigma} \;  d \theta \wedge d \varphi
				\, .
				\label{eq:StokesConvention}
			\end{eqnarray}

			We wish to analyze energy and angular momentum in a ball inside the cosmological horizon $\Hor$. The energy is a hamiltonian charge associated with the Killing vector
			\begin{equation}
				\KilT=\partial_{t} \, .
			\end{equation}
			For Angular momentum, the hamiltonian fields $\KilRx, \KilRy, \KilRz$ are given by \eqref{eq:KilRx}--\eqref{eq:KilRz}. Using \eqref{eq:hopftwoformrho}, we obtain the explicit form of the energy current \eqref{eq:Ecurrent} for hopfion-like solution
			\begin{align}
				\mcH^{t} [\KilT]&=
				\frac{ \sin \theta }{4 \pi \sinh^2(\alpha \rho)} \left\{ 4 \alpha^2 \sin^2 \theta  \chitrho  \chitrhobar +
				\left[\left((\cos^2 \theta + 1) \partial_{\rho} \chitrho - 2 \cos \theta \partial_{t} \chitrho\right) \partial_{\rho} \chitrhobar
				\right.
				\right.
				\nonumber
				\\
				&
				\left.
				\left.
				+\left(-2 \cos \theta \partial_{\rho} \chitrho +  (1 + \cos^2 \theta)\partial_{t} \chitrho\right)\partial_{t} \chitrhobar \right] \sinh^2(\alpha \rho)  \right\}
				\, .
			\end{align}
			We recall that $\chitrho$ is given by \eqref{eq:chitrho2} together with \eqref{eq:upsilonPhase}. $\chitrhobar$ is a complex conjugate of $\chitrho$. Analogically, components of angular momentum current read
			\begin{align}
				\mcH^{t} \left[\KilRx\right]&= 	
				\frac{\chitrho \sin^2 \theta \left[\left(\cos \theta \sin \varphi -\imath \cos \theta \cos \varphi \right) \partial_{t} \chitrhobar-\left(\sin \varphi-\imath   \cos^2 \theta \cos \varphi \right)\partial_{\rho} \chitrhobar \right] }{2 \pi} 
				\nonumber
				\\
				& +
				\frac{\chitrhobar \sin^2 \theta \left[(\sin \varphi+\imath \cos \varphi)\cos \theta \,  \partial_{t} \chitrho-\left(\sin \varphi+\imath \cos^2 \theta \cos \varphi\right)\, \partial_{\rho} \chitrho  \right]}{2 \pi}
				\, ,	
				\\
				\mcH^{t} \left[\KilRy\right]&=
				\frac{\chitrho \sin^2 \theta \left[ \left(\cos \varphi+\imath \cos^2 \theta \sin \varphi \right) \partial_{\rho} \chitrhobar-\left(\cos \theta \cos \varphi + \imath \sin \varphi \cos \theta \right) \partial_{t} \chitrhobar \right] }{2 \pi}
				\nonumber
				\\
				&
				+ 
				\frac{\chitrhobar \sin^2 \theta \left[\left(\cos \varphi-\imath \cos^2 \theta \sin \varphi\right)\partial_{\rho} \chitrho-(\cos \varphi-\imath \sin \varphi)\cos \theta \, \partial_{t} \chitrho \right]}{2 \pi}
				\, ,
				\\
				\mcH^{t} \left[\KilRz\right]&=
				\frac{\imath \sin^3 \theta \left[\chitrhobar  \left(\cos \theta \partial_{\rho} \chitrho - \partial_{t} \chitrho \right) - \chitrho \left(\cos \theta \partial_{\rho} \chitrhobar-\partial_{t} \chitrhobar\right)\right]}{2 \pi}
				\, .
			\end{align}
			As the next step, we calculate quasi-local densities of the above currents. By quasi-local density we mean an integral from current over sphere
			\begin{equation}
				\mathit{Q}[X]:=\int_{S(R)} \mcH^{t} [X] d \sigma \, .
			\end{equation} 
			The quasi-local density of energy reads
			\begin{align}
				\lefteqn{
					\mathit{Q}[\KilT]:=
					\int_{S(\rho(R))} \frac{ \sin \theta }{4 \pi \sinh^2(\alpha \rho)} \left\{ 4 \alpha^2 \chitrho  \chitrhobar \sin^2 \theta +
					\left[\left((1+\cos^2 \theta) \partial_{\rho} \chitrho - 2 \cos \theta \partial_{t} \chitrho\right) \partial_{\rho} \chitrhobar
					\right.
					\right.
				}&
				\nonumber
				\\
				&
				\left.
				\left.
				+\left((1 + \cos^2 \theta)\partial_{t} \chitrho-2 \cos \theta \partial_{\rho} \chitrho\right)\partial_{t} \chitrhobar \right] \sinh^2(\alpha \rho)  \right\}
				d \theta \wedge d \varphi
				\nonumber
				\\
				&= 
				\frac{\sinh^2(\alpha \rho) \partial_{\rho} \chitrho \, \partial_{\rho} \chitrhobar +\sinh^2(\alpha \rho) \partial_{t} \chitrho \, \partial_{t} \chitrhobar +2 \alpha^2 \chitrho \chitrhobar}{3 \sinh^2(\alpha \rho)}
				\nonumber
				\\
				&=
				\frac{2 \alpha^6 \sinh^2(\alpha \rho)\left[3(\cosh (\alpha \rho) \cosh (\alpha t)-\cos \upsilon)^2+\sinh^4(\alpha \rho)\right] }{3(1+ \alpha^2 \beta^2)^4\left[(\cosh (\alpha \rho)-\cosh (\alpha t) \cos \upsilon)^2+\sin^{2} \upsilon\sinh^2(\alpha \rho)\right]^3}
				\, .
				\phantom{xxxxxxxxxxxx}
			\end{align}
			Note that the quasi-local density of energy is manifestly non-negative. Considering the quasi-local density of angular momentum, it turns out that only one component is non-vanishing
			\begin{align}
				\mathit{Q}\left[\KilRz \right]&:=\int_{S(R)}
				\frac{\imath \sin^3 \theta \left[\chitrhobar  \left(\cos \theta \partial_{\rho} \chitrho - \partial_{t} \chitrho \right) - \chitrho \left(\cos \theta \partial_{\rho} \chitrhobar-\partial_{t} \chitrhobar\right)\right]}{2 \pi}
				d \theta \wedge d \varphi
				\nonumber
				\\
				&= 
				\frac{4 \imath \chitrho\left(\partial_{t} \chitrhobar \right)}{3}-\frac{4 \imath\left(\partial_{t} \chitrho\right) \chitrhobar}{3}
				\nonumber
				\\
				&=
				\frac{16 \alpha^5 \sinh (\alpha \rho)^4 \sin \upsilon(\cos \upsilon-\cosh (\alpha \rho) \cosh (\alpha t)) }{3(1+ \alpha^2 \beta^2)^4\left[(\cosh (\alpha \rho)-\cosh (\alpha t) \cos \upsilon)^2+\sin^{2} \upsilon\sinh^2(\alpha \rho)\right]^3}
				\, .
			\end{align}
			We are now ready to obtain the total energy inside the cosmological horizon
			\begin{align}
				\EdS&= \int_{0}^{\infty} \mathit{Q}[\KilT]   d \rho
				\nonumber
				\\
				&=\frac{2 \alpha(5+\cos (2 \upsilon)-6 \upsilon \cot \upsilon)}{3\left(1+\alpha^2 \beta^2\right)^4 \sin^4 \upsilon}
				\label{eq:EnergyHopfFinal}
			\end{align}
			Analogically, we find the total angular momentum
			\begin{align}
				\JdSz&=\int_{0}^{\infty} \mathit{Q}\Big[\KilRz\Big]   d \rho 
				\nonumber
				\\
				&=\frac{(48 \upsilon-32 \sin (2 \upsilon)+\sin(4 \upsilon))}{24\left(1+\alpha^2 \beta^2\right)^4 \sin^4 \upsilon}
			\end{align}
			Now, we wish to investigate fluxes of energy and angular momentum. To proceed, we use the formula \eqref{eq:FluxDef} where the Lie derivative is calculated with respect to
			\begin{equation*}
				Y=\KilT=\partial_{t} \, ,
			\end{equation*}  
			and integrate over $\Sigma_{t}$. Taking into account that the formula is a total divergence, we use the divergence theorem to find
			\begin{align}
				\frac{d}{d t}\int_{\Sigma_{t}} \mcH^{\mu} [X] dS_\mu
				&=
				\int_{\Sigma_{t}} \Lie_{\KilT} \mcH^{\mu} [X] dS_\mu
				\nonumber
				\\
				&=
				-\frac{1}{2 \pi} \int_{\Sigma_{t}} \sqrt{|\det g|} \nabla_{\sigma} \left[\KilT^{[\sigma} F^{\mu] \alpha}
				X^{\kappa}F_{\kappa \alpha} - \frac{1}{4} \KilT^{[\sigma} X^{\mu]} F^{\alpha \beta} F_{\alpha \beta} \right] dS_\mu 
				\nonumber
				\\
				&=
				\frac{1}{2 \pi} \int_{S(R)} \sqrt{|\det g|} \left[\KilT^{[\sigma} F^{\mu] \alpha}
				X^{\kappa}F_{\kappa \alpha} - \frac{1}{4} \KilT^{[\sigma} X^{\mu]} F^{\alpha \beta} F_{\alpha \beta} \right] dS_{\sigma \mu}
				\nonumber
				\\
				&=
				\frac{1}{4 \pi} \int_{S(\rho(R))} \sqrt{|\det g|} \left[ F^{\rho \alpha}
				X^{\kappa}F_{\kappa \alpha} - \frac{1}{4}  X^{\rho} F^{\alpha \beta} F_{\alpha \beta} \right] d \theta d \varphi
				\, ,
			\end{align}
			where the convention on orientation in Stokes theorem is given by \eqref{eq:StokesConvention}. Let us emphasize that a field with an index $\rho$ denotes a $\rho$-component of a field in $(t,\rho, \theta, \varphi)$ coordinates. The flow of energy reads
			\begin{align}
				\frac{d \mathit{E}}{d t}&=\frac{1}{4 \pi} \int_{S(\rho(R))} \sqrt{|\det g|} F^{\rho \alpha}
				\KilT^{\kappa}F_{\kappa \alpha} d \theta d \varphi
				\nonumber
				\\
				&=\frac{1}{4 \pi} \int_{S(\rho(R))} \sqrt{|\det g|} F^{\rho \alpha}
				F_{t \alpha} d \theta d \varphi
				\, ,
			\end{align}
			For hopfion-like solution \eqref{eq:hopftwoformrho}, we find
			\begin{align}
				\frac{d \mathit{E}}{d t}&= \frac{4\left(\partial_{t} \chitrho\right)\left(\partial_{\rho} \chitrhobar\right)}{3}+\frac{4\left(\partial_{\rho} \chitrho\right)\left(\partial_{t} \chitrhobar\right)}{3}
				\nonumber
				\\
				&=
				-
				\frac{32 \alpha^6 \sinh^3 (\alpha \rho) \sinh (\alpha t)[\cosh (\alpha \rho) \cosh (\alpha t)-\cos \upsilon] }{3(1+ \alpha^2 \beta^2)^4\left[(\cosh (\alpha \rho)-\cosh (\alpha t) \cos \upsilon)^2+\sin^{2} \upsilon\sinh^2(\alpha \rho)\right]^3}
				\, .
				\label{eq:EnergyFluxHopfFinal}
			\end{align}
			We note that the energy flux is explicitly negative. Evaluating the flux when $\rho$ tends to infinity, we find the energy flow through cosmological horizon. Irrespective of the sign of alpha, we obtain
			\begin{equation}
				\lim\limits_{\rho \to \infty} \frac{d \mathit{E}}{d t} =0 \, .
			\end{equation}
			The hamiltonian vector fields $\KilRx, \KilRy, \KilRz$, given by \eqref{eq:KilRx}-\eqref{eq:KilRz}, are tangent to spheres, so the fluxes of angular momentum yield
			\begin{align}
				\frac{d \! \JdSx}{d t}&=\frac{1}{4 \pi} \int_{S(\rho(R))} \sqrt{|\det g|} F^{\rho \alpha}
				\KilRx{}^{C}F_{C \alpha} d \theta d \varphi
				\, ,
				\\
				\frac{d \! \JdSy}{d t}&=\frac{1}{4 \pi} \int_{S(\rho(R))} \sqrt{|\det g|} F^{\rho \alpha}
				\KilRy{}^{C}F_{C \alpha} d \theta d \varphi
				\, ,
				\\
				\frac{d \! \JdSz}{d t}&=\frac{1}{4 \pi} \int_{S(\rho(R))} \sqrt{|\det g|} F^{\rho \alpha}
				\KilRz{}^{C}F_{C \alpha} d \theta d \varphi
				\, .
			\end{align}
			Evaluating the above functional on hopfion-like solution \eqref{eq:hopftwoformrho}, we obtain only non-vanishing projection of angluar momentum on $\KilRz$
			\begin{align}
				\label{eq:AngMomFlux}
				\frac{d \! \JdSz}{d t}&=\frac{4 \imath \chitrho \left(\partial_{\rho} \chitrhobar\right) }{3} -\frac{4 \imath \chitrhobar\left(\partial_{\rho} \chitrho\right)}{3}
				\nonumber
				\\
				&=
				\frac{16 \alpha^5 \sinh^5 (\alpha \rho)  \sinh (\alpha t) \sin \upsilon}{3(1+ \alpha^2 \beta^2)^4\left[(\cosh (\alpha \rho)-\cosh (\alpha t) \cos \upsilon)^2+\sin^{2} \upsilon\sinh^2(\alpha \rho)\right]^3}
				\, .
			\end{align}
			The sign of the flux of angular momentum depends on the sign of numerator in \eqref{eq:AngMomFlux}, which is determined on the combination of signs of $\alpha, t $ and $\upsilon$. $\upsilon$ is defined by \eqref{eq:upsilonPhase}. Calculating $\rho \to \infty$, we obtain flux of angular momentum through the cosmological horizon
			\begin{equation}
				\lim\limits_{\rho \to \infty} \frac{d \! \JdSz}{d t} = 0 \, .
			\end{equation}

    	\subsection{Topological charges associated with the duality symmetry. Helicity current \label{sec:TopoCharges}}

        	\noindent
        	The introduction to the section is a generalization of electromagnetic helicity to de Sitter space-time. Detailed results for Minkowski space-time can be found in Aghapour et al. \cite{aghapour2020helicity}. The Maxwell equations \eqref{eq:MaxwellEq1}-\eqref{eq:MaxwellEq2} remain manifestly invariant under the duality reflection $F_{\alpha\beta} \to (\ast F)_{\alpha\beta}$. However, the Lagrangian density \eqref{23V21.1} fails to be duality invariant. This motivates the introduction of a manifestly duality symmetric variation principle for the Maxwell equations.
        	
        	\noindent
        	We introduce an auxiliary two-form field, for which we will impose constraint
        	\begin{equation}
        		\label{eq:MaxwellTwoFormConstraint}
        		G_{\alpha\beta} = (\ast F) _{\alpha\beta}
        		\, .
        	\end{equation}
        	Maxwell equation $d G=0$ enables us to introduce a potential one-form
        	\begin{equation}
        		G_{\alpha\beta} = 2\,\partial_{[\alpha} C_{\beta]}\, .
        	\end{equation}
        	In \cite{cameron2012electric,bliokh2013dual}, the duality symmetric Lagrangian has been introduced for Minkowski space-time. Generalization of the Lagrangian to de Sitter space-time in our conventions yields
        	\begin{equation}\label{eq:dualLag}
        		\mcL_{dual} (\partial A_{\mu}, \partial C_{\nu} ) =- \frac {1}{32 \pi}  \sqrt{|\det g|}
        		\big( 
        		g^{\mu\nu} g^{\alpha\beta} F_{\mu \alpha} F_{\nu\beta}+g^{\mu\nu} g^{\alpha\beta} G_{\mu \alpha} G_{\nu\beta}
        		\big)
        		\,.
        	\end{equation}
        	with the corresponding action 
        	\begin{equation} \label{eq:ActionDual}
        		S_{dual} = \int d^4 r \, \mcL_{dual}
        	\end{equation} 
        	where we treat $C_\alpha$ and hence also $G_{\alpha\beta}$ as fields independent of $A_\alpha$ and $F_{\alpha\beta}$. The Lagrangian $\mcL_{dual}$ is invariant under the action of the $U(1)$ group
        	\begin{subequations}
        		\label{duality_tr_F}
        		\begin{equation}
        			F_{\alpha\beta}\to{} F_{\alpha\beta}\, \cos\lambda + G_{\alpha\beta}\, \sin\lambda \, ,		
        		\end{equation}
        		\begin{equation}
        			G_{\alpha\beta}\to{} G_{\alpha\beta}\, \cos\lambda - F_{\alpha\beta}\, \sin\lambda \, . 
        		\end{equation}
        	\end{subequations}
        	Recall that the constraint \eqref{eq:MaxwellTwoFormConstraint} holds. In addition to that, the dual Lagrangian \eqref{eq:dualLag} is invariant under, the duality transformation \eqref{duality_tr_F} in terms of potentials
        	\begin{subequations}
        		\begin{equation}
        			A\to  A\,\cos\lambda +  C\,\sin\lambda \,,
        		\end{equation}
        		\begin{equation}
        			C\to  C\,\cos\lambda -  A\,\sin\lambda \,.	 
        		\end{equation}
        	\end{subequations}  
        	This internal symmetry and leads to the conservation of helicity. The Noether current for the duality symmetry \eqref{duality_tr_F} is known as the helicity current

        	\begin{equation}
        		\label{eq:helicity_current}
        		\mathcal{C}^\alpha_{\mathcal H} = \frac{\partial \mcL_{dual}}{\partial (\partial_\alpha A_\beta)}\,C_\beta -\frac{\partial \mcL_{dual}}{\partial (\partial_\alpha C_\beta)}\,A_\beta 
        		= \frac{1}{8 \pi} \sqrt{|\det g|} (G^{\alpha\beta}\,A_\beta - F^{\alpha\beta}\,C_\beta), 
        	\end{equation}
        	for which the divergence vanishes 
        	\begin{equation}
        		\partial_{\alpha} \mathcal{C}^\alpha_{\mathcal H} =0 \, .
        	\end{equation}
        	This further implies that the flux equation is satisfied for \eqref{eq:helicity_current}, namely
        	\begin{align}
        		\Lie_{Y}\mathcal{C}^\alpha_{\mathcal H} &=\partial_{\sigma} \left(Y^{\sigma}\mathcal{C}^\alpha_{\mathcal H} \right) - \mathcal{C}^\sigma_{\mathcal H} \partial_{\sigma} Y^{\alpha} 
        		\nonumber
        		\\
        		&= 2 \partial_{\sigma} \left(Y^{[\sigma}\mathcal{C}^{\alpha]}_{\mathcal H} \right) +Y^{\alpha} \underbrace{\partial_{\sigma} \mathcal{C}^{\sigma}_{\mathcal H}}_{=0}
        		\, , 
        		\label{eq:ref_helicity_flux}
        	\end{align}
        	where in the first equality, we have used the definition of Lie derivative of vector density.
        	We distinguish the components of helicity current, the total helicity density
        	\begin{equation}
        		\mathcal H \equiv \mathcal{C}^t_{\mathcal H} \, ,
        	\end{equation}
        	and 
        	the helicity flux density also known as the spin density
        	\begin{equation}
        		\mathbf{S}=\mathbfcal{C}^{k}_{\mathcal H} \partial_{k} \, .
        	\end{equation}
        	The objects defined above are similar to well-known ones in optics. Restricting ourselves to Minkowski spacetime, in transverse gauge\footnote{By transverse gauge, we mean the following conditions
        		\begin{equation*}
        			A^0=C^0=0, \quad \nabla \cdot \mathbf{A}=\nabla \cdot \mathbf{C}=0 \, .
        		\end{equation*}
        		This gauge condition is a combination of the temporal gauge $A^0=C^0=0$, and the Coulomb gauge $\nabla \cdot \boldsymbol{A}=\nabla \cdot \boldsymbol{C}=0$.
        	}, the total helicity density reads
        	\begin{equation}
        		\mathcal H   = \frac{1}{8 \pi} \sqrt{|\det \eta|} \left(\mb A\cdot \mb B - \mb C \cdot \mb E\right),
        	\end{equation}
        	where $\eta$ is the metric of Minkowski space-time. The spin density is usually defined as
        	\begin{equation}
        		\mathbfcal{S}= \frac{1}{8 \pi} \sqrt{|\det \eta|} (\mb E \times \mb A + \mb B\times \mb C )\,.
        	\end{equation} 

        	\noindent
        	To proceed, we pass to our analysis of hopfion-like solution in de Sitter space-time. A convenient way to proceed is to develop the construction with anti-self dual Maxwell two-form \eqref{eq:AntiSelfDualDef} for hopfion-like solution. In this framework, the helicity current \eqref{eq:helicity_current} becomes
        	\begin{equation}
        		\mathcal{C}^\alpha_{\mathcal H} = \Im \left( \overline{\mathcal{F}}^{\alpha \beta}_{H} (\mathcal{A}_{H})_{\beta}
        		\right)
        	\end{equation} 
        	where $\overline{\mathcal{F}}_{H}$ denotes the complex conjugate of $\mathcal{F}_{H}$.
        	We continue with obtaining a potential one-form $\mathcal{A}$ for hopfion-like two-form \eqref{eq:hopftwoformrho}, which fulfills 
        	\begin{equation}
        		\mathcal{F}_{H}=d \mathcal{A}_{H} \, .
        	\end{equation}
        	The potential has a form
        	\begin{equation}
        		\begin{split}
        			\mathcal{A} = \frac{a I^2 e^{\imath \varphi}\sin{\theta}}{D_1 \left(1+a^2 b^2\right)^2}
        			\biggr\{\left[\imath\sin{v}\sinh{u} + (1+\cos{v})\left( \frac{\sinh(a r)}{I}-a^{-1}\partial_r(\cosh{u}) \right)\right]dt +\\ +\left[-\imath\sin{v}\cosh{u}+(1+\cos{v})\left( \frac{\sinh(a t)}{I}-a^{-1}\partial_t(\cosh{u}) \right)\right]dr\biggr\} +\\
        			\frac{a e^{\imath \varphi}\sinh(a r)}{D_2 \left(1+a^2 b^2\right)^2}
        			\biggr\{
        			(I(\cosh{u}-\cosh(a t-\imath v))-\cos{v}\sinh^2(a r))(\sin{\theta}d\varphi-\imath\cos{\theta}d\theta) + \\
        			\sinh(a r) (\cosh(a r)\sin{v} + \imath (\sinh(a t)+\sinh(at - \imath v)))(d\theta + \imath \sin{\theta}\cos{\theta}d\varphi)
        			\biggr\}
        		\end{split}
        		\label{eq:PotentialOneForm}
        	\end{equation}
        	where $I = \cosh(a r) + \cosh(a t)$,
        	\begin{equation}
        		\begin{split}
        			D_1 &=  (\cosh (a r)-\cosh (a t-i v)) \left(\sin (v) \cosh (a r)+2 i \cos \left(\frac{v}{2}\right) \sinh \left(a t-\frac{i v}{2}\right)\right)^2\\
        			D_2 &=  (\cosh (a r)-\cosh (a t-i v))^2 \left(\sin (v) \cosh (a r)+2 i \cos \left(\frac{v}{2}\right) \sinh \left(a t-\frac{i v}{2}\right)\right)
        		\end{split}
        	\end{equation}
        	and $u = u(t,r)$ is taken, such that
        	\begin{equation}
        		\begin{split}
        			\cosh{u} &= \frac{1 +\cosh (a r) \cosh (a t)}{\cosh (a r)+\cosh (a t)} \, , \\
        			\sinh{u} &= \frac{\sinh (a r) \sinh (a t)}{\cosh (a r)+\cosh (a t)} \, .
        		\end{split}
        	\end{equation}
        	\normalsize
        	Using \eqref{eq:PotentialOneForm} and \eqref{eq:hopftwoformrho}, we obtain the total helicity for hopfion-like solution
        	\begin{align}
        		H&=\int_{\mathcal{B}_{\Hor}} \mathcal H dS_t
        		\nonumber
        		\\
        		&=\frac{1}{8 \pi}\Im \int_{0}^{\infty} d \rho \int_{0}^{\pi} d \theta \int_{0}^{2 \pi} d \varphi  \sqrt{|\det g|} \overline{\mathcal{F}}^{t \alpha}_{H} (\mathcal{A}_{H})_{\alpha}
        		\nonumber
        		\\
        		&=
        		\frac{\left(\alpha^4 b^4 \arctan \left(\frac{1}{\alpha b}\right)+2 b^2 \alpha^2 \arctan \left(\frac{1}{\alpha b}\right)-b^3 \alpha^3+\arctan \left(\frac{1}{\alpha b}\right)+\alpha b\right)}{8 b^4\left(b^2 \alpha^2+1\right)^2}
        		\, .
        		\label{eq:HelicityHopfionLike}
        	\end{align}
        	Note that the helicity current is time independent. When cosmological constant tends to zero, we find
        	\begin{equation}
        		H_{\alpha \to 0}= \frac{ \pi \operatorname{signum}(b)}{16 b^4}+\mathrm{O}\left(\alpha^3\right) \, .
        	\end{equation} 
        	
        	Analogically, with the use of anti-self dual Maxwell two-form, the spin density takes the form
        	\begin{equation}
        		\mathbfcal{S}= \frac{1}{8 \pi} \sqrt{|\det g|} \Im \left( \overline{\mathcal{F}}^{k \alpha}_{H} (\mathcal{A}_{H})_{\alpha}
        		\right)
        		\partial_{k}
        		\, .
        	\end{equation}
        	Using $Y=\KilT=\partial_{t}$ for \eqref{eq:ref_helicity_flux}, we obtain the helicity flux formula through cosmological horizon
        	\begin{align}
        		\frac{d H}{d t}&=\int_{\mathcal{B}_{\Hor}} \Lie_{\KilT}\mathcal{C}^{\mu}_{\mathcal H} dS_\mu
        		\nonumber
        		\\
        		&=2 \int_{\mathcal{B}_{\Hor}}  \partial_{\sigma} \left(Y^{[\sigma}\mathcal{C}^{\mu]}_{\mathcal H} \right) dS_\mu
        		\nonumber
        		\\
        		&=-\int_{\Hor} \left(Y^{\sigma}\mathcal{C}^{\mu}_{\mathcal H} \right)dS_{\sigma \mu}
        		\nonumber
        		\\	
        		&=-\frac{1}{8 \pi} \int_{\Hor} \sqrt{|\det g|} \Im \left( \overline{\mathcal{F}}^{\rho \alpha}_{H} (\mathcal{A}_{H})_{\alpha}
        		\right) d \theta d \varphi
        		\nonumber
        		\\
        		&=0 \, . 
        	\end{align} 

    \section{Conclusions}
    		The paper presents a study that expands on previous work on Hopfion solutions in flat space-time and extends it to the space-time of de Sitter geometry. The study introduces new electromagnetic solutions with non-trivial topological properties constructed using reduced data. A one-parameter family of such solutions has been found, which becomes electromagnetic Hopfions in flat space-time as the cosmological constant $\Lambda$ tends to zero. The solutions have been obtained with the help of conformal correspondence between Minkowski space-time and de Sitter space-time. The origin of other issues resolved in the section is that the global structures of de Sitter and Minkowski space-times are incompatible. In addition, the inversion-like operation for the solution parameter $\beta$, see \eqref{eq:beta_param}, is required to obtain generalized hopfions when the cosmological constant tends to zero. 
    		Appendix \ref{sec:ReducedDataDeSitter} describes the recovery procedure that enables us to obtain the Maxwell two-form from the reduced data, together with the necessary mathematical background from appendix \ref{sec:math_sup}.
    		
    		In the following parts of the paper, we focus our analysis on a member of such a family, which becomes classical Ra\~nada's Hopfion in the limit of Minkowski space. The Maxwell two-form for this solution is given by \eqref{eq:FHtr}. In chapter \ref{sec:TopoProperties}, we investigate the basic topological properties of the solution \eqref{eq:FHtr}, including verifying that the field lines are tangent to the spherical foliation on $\{t=0\}$ surface. The integral curves of the electromagnetic fields have a structure of linked circles, entirely analogical to the Hopf fibration (described at the beginning of section \ref{sec:TopoProperties}). 
    		Using Hamiltonian methods in section \ref{sec:KillingCharges}, we analyze the energy, angular momentum, and topological charges for the constructed solution. The energy \eqref{eq:EnergyHopfFinal} inside a ball surrounded by the cosmological horizon for a Hopfion-like solution remains finite and constant in time. Its flux \eqref{eq:EnergyFluxHopfFinal} at any time $\{t=const.\}$ vanishes on the cosmological horizon. Similar results are obtained for angular momentum.
    		
    		In section \ref{sec:TopoCharges}, we have generalized Noether flux of duality symmetry, known as helicity current, for de Sitter space-time. Then, we obtained the helicity current for the hopfion-like solution in de Sitter; see the equation \eqref{eq:HelicityHopfionLike} and comments below.

    	\subsection{Perspective of further research}
    	Generalized hopfion-like solution on de Sitter background has potential for further investigations, namely:
    	\begin{enumerate}
    		\item \textbf{Analysis outside the horizon, in particular on the conformal boundary.}\\
    		The topological properties of electromagnetic fields in de Sitter spacetime is poorly known, especially outside the horizon. We plan to perform a detailed analysis of electromagnetic fields, in particular hopfion-like solutions, in the regime of conformal infinity $\mathcal{I}^{+}$.   
    		
    		\item \textbf{Hopfion-like solution for weak gravitational field on de Sitter background.}\\
    		The reduced data for electromagnetism \eqref{eq:reduced_data_tb}, viewed as a scalar field, is simultaneously a reduced data for weak gravitational waves for de Sitter background.

    		In the case of Kottler background, the construction is given in \cite{jezierski2021gauge}. A dedicated analysis for wave solutions on the de Sitter background is a generalization of appendix C.3 in \cite{Smolka_2018}. We highlight that the reduced data
    		\begin{equation}
    			\Psi_{l}=\Phi_l
    			\, ,
    		\end{equation}
    		viewed \eqref{eq:reduced_data_tb} as a reduced data for linearized gravity. The linearized gravity solution has analogical properties to EM solution. It will be analyzed in details in separate paper. 
    		
    	\end{enumerate}

	\appendix
	\section{Scalar representation of Maxwell fields in de Sitter spacetime\label{sec:ReducedDataDeSitter}}

	The description of electromagnetism in terms of the complex scalar field has been introduced\footnote{The description has been developed for electromagnetism and linearized gravity. In the case of linearized gravity, the introductory paper is \cite{jezierski1995relation}.} by one of us in \cite{jezierski2002cyk}. In general, the description of electromagnetism in terms of complex scalar field can be generalized for type D spacetimes (Petrov classification). The construction,  based on conformal Yano--Killing tensor (see appendix \ref{sec:ElCYK}), is presented in detail in \cite{jezsmo}, together with a description of electromagnetism on Kerr background as an example.
	
	We begin the presentation of reduced data with a brief recall of the self-dual Maxwell field. The self-dual Maxwell field is a complex two-form that is related to the real Maxwell field as follows
	\begin{equation}
		\mathcal{F}_{\mu\nu} = F_{\mu\nu} - \imath (\ast F_{\mu\nu})
		\, ,
		\label{eq:SelfDualFDef}
	\end{equation}
	where $\imath^{2}=-1$\, , and $F_{\mu \nu}$ and $\ast F_{\mu \nu}$ are Maxwell-field and its Hodge dual companion respectively.
	The self-duality in Hodge-star\footnote{
		In the space of differential forms on an oriented manifold, one can define the Hodge duality (Hodge-star) mapping. It assigns to every $p$-form an $(n-p)$-form (where $n$ is the manifold's dimension). We consider the case of $n=4$ and $p=2$. The Hodge star then becomes a mapping which assigns to a two-form $\omega$ a two-form $\ast \omega$.
		We can express this mapping in the following way:
		\begin{equation}
			\ast \omega_{\alpha \beta}=\frac{1}{2}{\varepsilon_{\alpha \beta}}^{\mu \nu}\omega_{\mu \nu} \, , \label{Hodge}
		\end{equation}
		where $\varepsilon_{\alpha \beta \gamma \delta}$ is the antisymmetric Levi--Civita tensor determining the orientation of the manifold ($\frac{1}{4!} \varepsilon_{\alpha \beta \gamma \delta} \mathrm{d} x^{\alpha} \wedge  \mathrm{d} x^{\beta} \wedge \mathrm{d} x^{\gamma} \wedge \mathrm{d} x^{\delta}$ is the volume form of the manifold $M$). We have $\ast \ast \omega=- \omega$ for the Lorentzian metric.
		
	} sense reads
	\begin{equation}\label{eq:selfdual}
		\mathcal{F} = \imath( \ast \mathcal{F}) \, .
	\end{equation}
	Vacuum Maxwell equations are simply an external derivative of $\mathcal{F}$,
	\begin{equation}
		d \mathcal{F} = 0 \, ,
	\end{equation}
	which is equivalent to:
	\begin{equation}\label{eq:expanded_maxwell}
		\mathcal{F}_{\mu\nu,\gamma} + \mathcal{F}_{\nu\gamma,\mu} + \mathcal{F}_{\gamma\mu,\nu} = 0 \, .
	\end{equation} 
	In the paper, we use reduced data, which is geometrically defined as  
	\begin{equation}
		\Phi=\mathcal{F}_{\mu \nu} Q^{\mu \nu} \, ,
	\end{equation}   
	where $Q_{\mu \nu}$ denotes conformal Yano--Kiling tensor (two-form). In general, the method can be used for the fields with sources. However, the paper is devoted to radiational solutions. We restrict the presentation to vacuum, sourceless case.\\ 
	In general, skew-symmetric tensor (two-form) $Q_{\mu\nu}$ is the \emph{conformal Yano--Killing tensor} (or CYK tensor) for the metric $g_{\mu\nu}$ if it satisfies the equality:  
	\begin{equation}
		Q_{\lambda\mu;\nu} + Q_{\nu\mu;\lambda} = 
		\frac{2}{3}\left( 
		g_{\nu\lambda} Q^{\rho}_{\ \mu;\rho} + 
		g_{\mu(\lambda} Q_{\nu)}\! ^{\rho}_{\ ;\rho} 
		\right) \, ,
	\end{equation}
	De Sitter spacetime is a maximally symmetric space, containing a twenty-dimensional space of CYK tensor solutions, see \cite{jezierski2008conformal}\footnote{The paper \cite{jezierski2008conformal} is mainly devoted to anti-de Sitter. However, the considered proof in section 3 is valid for both de Sitter and anti-de Sitter spacetimes. }.
	The recovery procedure is based on a particular choice of the CYK solution
	\begin{equation}
		\widetilde{Q}_{\mu \nu} d x^{\mu} \wedge d x^{\nu} = r dt\wedge dr \, ,
	\end{equation}  
	for which the reduced data is proportional to the time-radius component of the self-dual Maxwell field
	\begin{equation}
		\Phi = \widetilde{Q}_{\mu\nu}\mathcal{F}^{\mu\nu} = r\mathcal{F}^{tr} = -r\mathcal{F}_{tr}
		\, .
	\end{equation}
	Recall that $(t,r,\theta,\varphi)$ are static coordinates. Vacuum Maxwell equations imply that $\Phi$ satisfies the generalized wave equation
	\begin{equation}
		\big(\Box - 2 \alpha^2 \big) \Phi=0 \, ,
		\label{eq:deSitterwave}
	\end{equation} 
	and vice-versa: in fact, any smooth solution of equation \eqref{eq:deSitterwave} is equivalent to a sourceless Maxwell field via the recovery procedure presented below. See appendix \ref{sec:ElCYK} for justification, in particular the equation \eqref{eq:DeSitterWave} and comments nearby. The construction is coordinate independent --- it requires only the existence of two-dimensional foliation of spacetime by spheres. For simplicity, we present the method in static coordinates $(t,r,\theta, \varphi)$. We show that six independent components of the self-dual Maxwell field can be determined by reduced data $\Phi$ and its derivatives.
	
	As the next step, we perform (2+2)-decomposition of the Maxwell field. The two-form is thus decomposed in terms of the spherical (angular) components and terms that belong to the normal (time-radius) bundle. The time-radius term is algebraically related to the reduced data
	\begin{equation}
		\mathcal{F}_{tr}=-\frac{\Phi}{r}
		\, .
	\end{equation}
	The self-dual property \eqref{eq:selfdual} enables us to obtain the angular part immediately
	\begin{equation}
		\mathcal{F}_{AB} = \imath( \ast \mathcal{F})_{AB} = \frac{1}{2}\imath \varepsilon_{AB\mu\nu}\mathcal{F}^{\mu\nu} = -\imath \varepsilon_{AB} \mathcal{F}_{tr} = \imath \varepsilon_{AB}\frac{\Phi}{r}
		\, ,
	\end{equation}
	where $\varepsilon_{AB}=\varepsilon_{t r A B}$. The Hodge--Kodaira decomposition\footnote{
		In general, Hodge--Kodaira decomposition states that each differential form, say $\omega$, on a closed Riemannian manifold can be uniquely decomposed as a sum of three parts in the form
		\begin{equation}
			\omega =d\alpha +\delta \beta +\gamma \, ,
		\end{equation}
		where $\delta$ is a co-derivative (the adjoint operator of $d$) and $\gamma$ is harmonic: $\Delta \gamma=0$.  
	} on the unit sphere is essential in the recovery procedure of the mixed terms. The mixed components, $\mathcal{F}_{tA}$ and $\mathcal{F}_{rA}$, are one-forms from the point of view of the cotangent bundle on the sphere. Considering that there are no non-trivial harmonic one-forms on the sphere, each one-form on the sphere can be decomposed as a sum of a differential and a co-differential of certain complex functions. For $\mathcal{F}_{tA}$, we have
	\begin{equation}
		\mathcal{F}_{tA} = \xi_{,A} + \varepsilon_A^{\ B} \beta_{,B}
		\, .
	\end{equation}
	The self-duality property \eqref{eq:selfdual} leads to
	\begin{equation}
		\mathcal{F}_{rA} 
		= \imath ( \ast \mathcal{F})_{tA} 
		= \imath \varepsilon_{rAtB} \mathcal{F}^{tB} 
		= 
		\frac{\imath}{1-\alpha^2r^2} \varepsilon_A^{\ B} \mathcal{F}_{tB} 
		= \frac{\imath}{1-\alpha^2r^2} 
		\left( -\beta_{,A} + \varepsilon_A^{\ B} \xi_{,B} \right)
		\, .
	\end{equation}
	The Maxwell equations \eqref{eq:expanded_maxwell} allow us to obtain formulas for $\Delta\xi$ and $\Delta\beta$, where $\Delta$ is a two-dimensional laplacian on the unit sphere. In particular, we obtain: 
	\begin{equation}
		\begin{split}
			\mathbf{\Delta}\xi &= (1-\alpha^2r^2)\partial_r (r\Phi) \\
			\mathbf{\Delta} \beta &= i\partial_t (r\Phi)
		\end{split}
	\end{equation}
	The spherical Laplace operator can be quasi-locally inverted using the methods presented in \ref{sec:math_sup}. Thus, the Maxwell tensor can be reconstructed using the reduced data scalar function $\Phi$.
	
	\section{Mathematical supplement \label{sec:math_sup}}
		\subsection{Three-dimensional Laplace operator and its inverse}
		Consider Laplace equation
		\begin{equation}
			\triangle G(\mathbf{r},\mathbf{r}')= - \delta^{(3)}(\mathbf{r}-\mathbf{r}') \label{eq:3_D_laplace}
		\end{equation}
		with a solution on an open set without a boundary. $\delta^{(3)}(\mathbf{r}-\mathbf{r}')$ is a three-dimensional Dirac delta. $G(\mathbf{r},\mathbf{r}')$ is the following Green function of (\ref{eq:3_D_laplace}):
		\begin{equation}
			G(\mathbf{r},\mathbf{r}')=\frac{1}{4 \pi ||\mathbf{r}-\mathbf{r}'||}
			\, .
		\end{equation}
		The solution of the Poisson equation
		\begin{equation}
			\triangle u(\mathbf{r})=-f(\mathbf{r})
		\end{equation}
		is the convolution of $f(\mathbf{r})$ and Green function
		\begin{equation}
			u(\mathbf{r})=\int_{\Sigma} f(\mathbf{r}') G(\mathbf{r},\mathbf{r}') \mathrm{d} \mathbf{r}'=\int_{\Sigma} \frac{f(\mathbf{r}')}{4 \pi ||\mathbf{r}-\mathbf{r}'||} \mathrm{d} \mathbf{r}'
			\, .
		\end{equation}
		\subsection{Two-dimensional Laplace operator and its inverse}
		Consider a two-dimensional unit sphere in $\mathbb{R}^3$, parameterized by a unit position vector $\mathbf{n}$. One of the main differences is that the domain of the solutions is the compact surface without boundary. The conclusions of the Stokes theorem $(\int_{\mathbb{S}^2} \mathbf{\Delta} u(\mathbf{n})=0)$ require a modified problem to be examined than in the three-dimensional case. Consider the following two-dimensional Laplace equation with an additional condition
		\begin{eqnarray}
			\mathbf{\Delta} \mathbf{G}(\mathbf{n},\mathbf{n'})&=&1- \delta^{(2)}(\mathbf{n}-\mathbf{n'}) \, , \label{eq:2_D_laplace} \\
			\int_{\mathbb{S}^2} \sigma \mathbf{G}(\mathbf{n},\mathbf{n'}) \mathrm{d} \mathbf{n'}&=&0 \, ,
		\end{eqnarray}
		where $\sigma$ is area element on $\mathbb{S}^2$. We have the solution
		\begin{equation}
			\mathbf{G}(\mathbf{n},\mathbf{n'})=-\frac{1}{4 \pi}\left(\ln \left(\frac{1-\mathbf{n} \cdot \mathbf{n'}}{2} \right)+1\right) \, ,
		\end{equation}
		where `$\cdot$' is a scalar product of the position vectors\footnote{For a given point $(\theta, \varphi)$ in spherical coordinates on the unit sphere, the three-dimensional position vector in the Cartesian embedding is $n=\sin \theta \cos \varphi \partial_{x}+ \sin \theta \sin \varphi \partial_{y}+\cos \theta \partial_{z}$. Then, we use the scalar product with Euclidean metric.}. The solution of the Poisson equation
		\begin{eqnarray}
			\mathbf{\Delta} s(\mathbf{n})&=&-f(\mathbf{n}) \, ,  \\
			\int_{\mathbb{S}^2} \sigma f(\mathbf{n}) \mathrm{d} \mathbf{n}&=&0 \, ,
		\end{eqnarray}
		is the convolution of $f(\mathbf{n'})$ and the Green function
		\begin{equation}
			u(\mathbf{n})=\int_{\mathbb{S}^{2}} f(\mathbf{n'}) \mathbf{G}(\mathbf{n},\mathbf{n'}) \mathrm{d} \mathbf{n'}=-\int_{\mathbb{S}^{2}}\frac{1}{4 \pi}\left(\ln(\frac{1-\mathbf{n} \cdot \mathbf{n'}}{2})+1\right)  f(\mathbf{n'}) \mathrm{d} \mathbf{n'}
			\, .
		\end{equation}
		See \cite{CJM1998Uni} and \cite{jezierski2002peeling} for a detailed view, \cite{szmytkowski2006closed} is a specialized literature on the subject.
		\subsection{Operations on the sphere \label{sec:operation_on_S2}}
		Let us denote by $\mathbf{\Delta}$ the Laplace--Beltrami operator associated with the standard metric $h_{AB}$ on $\mathbb{S}^2$. Let $SH^l$ denote the space of
		spherical harmonics of degree $l$ ($g\in SH^l \Longleftrightarrow
		{\mathbf{\Delta}}g= -l(l+1)g$).
		Consider the following sequence
		\[
		\begin{array}{ccccccccc}
			V^0\oplus V^0 & \stackrel{i_{01}}{\longrightarrow} & V^1
			& \stackrel{i_{12}}{\longrightarrow} & V^2
			& \stackrel{i_{21}}{\longrightarrow} & V^1
			& \stackrel{i_{10}}{\longrightarrow} & V^0 \oplus V^0 \, .
		\end{array}
		\]
		Here $V^0$ is the space of, say, smooth functions on $\mathbb{S}^2$,
		$V^1$ -- that of smooth covectors on $\mathbb{S}^2$, and
		$V^2$ -- that of symmetric traceless tensors on $\mathbb{S}^2$.
		The various mappings above are defined as follows:
		\[ i_{01}(f,g)=f_{||a}+\varepsilon_a{^b}g_{||b} \, , \]
		\[ i_{12}(v)= v_{a||b}+ v_{b||a}-h_{ab}v^c_{||c} \, , \]
		\[ i_{21}(\chi)= \chi_a{^b}{_{||b}} \, , \]
		\[ i_{10}(v)=\left( {v^a}_{||a}, \varepsilon^{ab}v_{a||b} \right) \, , \]
		where $||$ is used to denote the covariant derivative with respect to
		the Levi--Civita connection of the standard metric $h_{AB}$ on
		$\mathbb{S}^2$. For more details, see Appendix E in \cite{jezierski2002peeling}.
		\subsection{Identities on the sphere \label{sec:spher_ident}}
		We have used the following identities on a sphere
		\begin{equation}
			-\int_{S(r)} \pi^A v_A = \int_{S(r)} (r{\pi}^{A}{_{|| A}}) {\mathbf{\Delta}} ^{-1}
			(rv^{A}{_{|| A}})+ \int_{S(r)} (r{\pi}^{A|| B}\varepsilon_{AB}) {\mathbf{\Delta}} ^{-1}
			(rv_{A|| B}\varepsilon^{AB})
		\end{equation}
		and similarly, for the traceless tensors, we have
		\begin{eqnarray}
			\int_{S(r)} \stackrel{\circ}{\pi}\! ^{AB} \stackrel{\circ}{v}\! _{AB} &=&
			2 \int_{S(r)} (r^2\varepsilon^{AC}
			\stackrel{\circ}{\pi}\! {_A{^B}}{_{||BC}})
			{\mathbf{\Delta}} ^{-1}({\mathbf{\Delta}}+2) ^{-1}
			(r^2\varepsilon^{AC}\stackrel{\circ}{v}\! {_A{^B}}{_{|| BC}})  \nonumber \\
			& & +\, 2 \int_{S(r)} (r^2\stackrel{\circ}{\pi}\! {^{AB}}{_{|| AB}})
			{\mathbf{\Delta}} ^{-1}({\mathbf{\Delta}}+2) ^{-1}
			(r^2\stackrel{\circ}{v}\! {^{AB}}{_{|| AB}}) \, .
		\end{eqnarray}
	\section{Electromagnetism and Yano--Killing tensors\label{sec:ElCYK}}
		For the purposes of this appendix, we consider an arbitrary four-dimensional Lorentzian manifold equipped with the metric $g$. The Levi-Civita derivative associated with the metric is denoted by `;'. We present a review of classical results about CYK two-forms in the context of electromagnetism. The section is based on \cite{jezierski_łukasik_2006} and classical references within.
		
		\noindent
		Conformal Yano--Killing two-form is a generalization of the Killing vector field to rank to anti-symmetric tensor.
		\begin{df}
			$Q_{\mu\nu}$ is a conformal Yano--Killing tensor if
			it fulfills the following equation:
			\begin{equation}\label{CYK_eq2}
				Q_{\lambda \kappa ;\sigma} +Q_{\sigma \kappa ;\lambda} =
				\frac{2}{3} \left( g_{\sigma \lambda}Q^{\nu}{_{\kappa ;\nu}} +
				g_{\kappa (\lambda } Q_{\sigma)}{^{\mu}}{_{ ;\mu}} \right) \, ,
			\end{equation}
			(first proposed by Tachibana and Kashiwada, cf. \cite{Tachibana1,Tachi-Kashi}).
		\end{df}
		
		\noindent
		In \cite{jezierski_łukasik_2006}, the duality property has been proven 
		\begin{thm}
			Let $g_{\mu\nu}$ be a metric tensor on a four-dimensional differential manifold $M$. An anti-symmetric tensor $Q_{\mu \nu}$ is a CYK tensor of the metric $g_{\mu\nu}$ if and only if its dual $\ast Q_{\mu \nu}$ is also a CYK tensor of this metric. \label{dualCYK}	
		\end{thm}
		
		\noindent
		The theorem above implies that solutions of Eq. (\ref{CYK_eq2}) exist in pairs for every four-dimensional manifold. Each solution can be assigned a dual solution (in the Hodge duality sense).
		
		\noindent
		In the context of electromagnetism, the crucial result in \cite{jezsmo} reads 
		\begin{thm} Let $F_{\mu\nu}$, $Q_{\mu\nu}$, ${C^{\sigma}}_{\lambda \nu \mu}$ and $R$ be respectively a Maxwell field, a CYK tensor, the Weyl tensor and the curvature scalar corresponding to the metric $g_{\mu\nu}$. Then
			\begin{equation}
				\left(\Box-\frac{1}{6}R \right)(F_{\mu \nu}Q^{\mu \nu})+\frac{1}{2}F^{\sigma \lambda}C_{\sigma \lambda \mu \nu}Q^{\mu \nu}=0 
				\, .
				\label{FQ_Weyl}
			\end{equation}
		\end{thm}
		\noindent
		The equation (\ref{FQ_Weyl}) remains unchanged under conformal transformation of the metric: ${g} \to
		\widetilde{g}=\Omega^2 g\,$ in view of the following property of CYK tensor 
		\begin{sthm}
			If $Q_{\mu \nu}$ is a CYK tensor for the metric $g_{\mu \nu}$, then $\Omega^{3} Q_{\mu\nu}$ is a CYK tensor for the conformally rescaled metric  $\Omega^{2} g_{\mu \nu}$.
		\end{sthm}
		\noindent
		We continue with the last puzzle in constructing the scalar wave equation for the evolution of Maxwell's theory--diagonalization of Weyl endomorphism. From now on, we have to restrict ourselves to the family of Pleba\'nski--Demia\'nski (P--D) generalized black hole space-times\cite{plebanski1976PD_metric,podolsky2006_PD_instr}. For the P--D family of space-times, Kubiz\v n\'ak and Krtou\v{s} \cite{kubizvnak2007CYK_PD} explicitly demonstrated that exist a pair\footnote{By pair, we mean two linearly independent two-forms which fulfill CYK equation \eqref{CYK_eq2}. According to the theorem \ref{dualCYK}, the solutions are dual of each other.} of CYK two-forms. we will denote them by $Y$ and $\ast Y$.

		Weyl curvature tensor has two pairs of anti-symmetric indices. The algebraic structure of the Weyl tensor
		$C_{\mu \nu}{^{\lambda \kappa}}$ allows it to be treated as an endomorphism in the space of two-forms at each point $p\in M$:
		\[ C:\operatornamewithlimits{\bigwedge}^{2} \mathrm{T}^{*}_p{M} \rightarrow \operatornamewithlimits{\bigwedge}^{2} \mathrm{T}^{*}_p{M}
		\, . \]
		In the six-dimensional space $\displaystyle\operatornamewithlimits{\bigwedge}^{2} \mathrm{T}^{*}_p{\mathrm{M}}$ we can distinguish a two-dimensional subspace $\mathbb{V}$. $\mathbb{V}$ is an invariant subspace of the endomorphism $C$. Surprisingly enough, $\mathbb{V}$ is spanned by $Y$ and $\ast Y$. More precisely,
		\begin{eqnarray}
			C_{\mu \nu}{^{\lambda \kappa}} \left(Y_{\lambda \kappa} + \imath \ast Y_{\lambda \kappa} \right)    =2 V \left( Y_{\mu \nu} + \imath \ast\! Y_{\mu \nu} \right) 
			\, ,
			\label{diag_Y}
		\end{eqnarray}
		where $ C_{\mu \nu}{^{\lambda \kappa}}$ is the Weyl tensor for the generalized black hole metric. The equation \eqref{FQ_Weyl} together with \eqref{diag_Y} enables us to formulate the following theorem:   
		\begin{thm}
			\label{thm:FIPDThm}
			Dynamics of a Maxwell field in the Pleba\'nski--Demia\'nski space-time can be reduced to the scalar wave equation:
			\begin{equation}
				\left(\Box -\frac{1}{6} R \right) \Phi + V \Phi=0 \, ,
				\label{eq:F_I_P_D}
			\end{equation}
			where $\displaystyle \Phi=\frac{\imath}{2} F^{\mu \nu} \left[Y_{\mu \nu}+ \imath(\ast Y_{\mu \nu})\right]$, R is the curvature scalar and $V$ is a complex scalar potential.
		\end{thm}
		\noindent
		The equation (\ref{eq:F_I_P_D}) has a complex potential, so it can not be treated as two independent real equations.
		
		If we restrict ourselves to the Kerr space-time, the equation (\ref{eq:F_I_P_D}) reduces to the one that was discovered for the first time by Fackerell and Ipser \cite{fackerell1972weak} in 1972. For this reason, we call the equation \eqref{eq:F_I_P_D} generalized Fackerell--Ipser equation. For Kerr space-time, it has the form
		\begin{equation}
			\widetilde{\Box} \Phi + \frac{2 m}{(r- \imath a \cos \theta)^3} \Phi=0 \, ,
		\end{equation}
		where $\widetilde{\Box}$ is d'Alembert operator for Kerr metric. $m$ and $m a$ are, respectively, the Kerr black hole's mass and angular momentum. $r$ and $\theta$ belong to Boyer--Lindquist coordinates. See \cite{jezsmo,TSPhD} for detailed results and discussion. With the help of the F--I equation, Andersson and Blue \cite{andersson2015uniform} have proven the boundedness of a positive definite energy of electromagnetic field on each hypersurface of constant time for a slowly rotated Kerr black hole.
		
		In the case of de Sitter space, the Weyl tensor is equal to zero due to the conformal flatness of this space. As a result, $V$ becomes zero in equation \eqref{eq:F_I_P_D}. The curvature scalar equals $12\alpha^2$. It allows us to identify the vacuum electromagnetic solutions in de Sitter space-time with the solutions of the generalized wave equation:
		\begin{equation}
			\label{eq:DeSitterWave}
			\left(\nabla_\lambda \nabla^\lambda - 2\alpha^2 \right)
			\Phi  = 0 \, .
		\end{equation} 

	\section{Killing vectors in de Sitter spacetime \label{sec:KillingVF}}
		For the de Sitter metric in the static form
		\begin{equation}\label{23IV22.1}
			g =   -(1-\alpha^2 r^2)dt^2 + \frac{dr^2}{1-\alpha^2 r^2}
			+ r^2
			(d\theta^2+\sin^2 \theta d\phi^2)
			\,,
		\end{equation}
		we have the following basis of the space of Killing vectors  in de Sitter spacetime
		\begin{eqnarray}
			\mathcal{T}&=&\partial_{t} \, ,
			\label{24VI21.t1}
			\\
			\mathcal{R} &=& \varepsilon^{B A} \zspaceD_{A}\big(R_{i} \wtx ^i \big) \partial_{B}\, ,
			\\
			\LKmom&=&\sqrt{\frac{1-\alpha r}{\alpha r+1}}e^{\alpha t}\Big[\LKmcon_{i} \wtx ^i \partial_{t}-\big(\alpha r +1\big)\LKmcon_{i} \wtx ^i \partial_{r}-\frac{\alpha r +1}{r} \zspaceD^{A}(\LKmcon_{i} \wtx ^i) \partial_{A} \Big]\, ,
			\label{24VI21.t2a}
			\\
			\LKbst&=&\sqrt{\frac{1+\alpha r}{1-\alpha r}}e^{-\alpha t}\Big[\LKbcon_{i} \wtx ^i \partial_{t}+\big(\alpha r -1\big)\LKbcon_{i} \wtx ^i \partial_{r}+\frac{\alpha r -1}{r} \zspaceD^{A}(\LKbcon_{i} \wtx ^i) \partial_{A} \Big]\, ,\phantom{xx}
			\label{24VI21.t2}
		\end{eqnarray}
		where
		$R_{i},\LKmcon_{i}$ and $\LKbcon_{i}$ are constants. The fields
		\begin{equation}
			\wtx ^{i}:=\frac{x^{i}}{r}
		\end{equation}
		are defined by \eqref{eq:nfields_def}. See also comments below. Detailed analysis of conformal Killing vectors is presented in \cite{jezczajka2018}.
	\section{Uniqueness of de Sitter conformally flat form \label{sec:UniqnessDesitter}}

		The de Sitter spacetime is conformally flat, while the Minkowski metric remains invariant under the actions of the Poincar\'e group, in contrast to the conformal factor. This raises the question of how rigid the choice of conformal factor is. We aim to demonstrate that the conformal transformation between de Sitter spacetime and Minkowski spacetime is not unique, even under quite strong assumptions. In order to do this, we consider coordinate transformations that preserve the spherical foliation consistent with the symmetry of the de Sitter metric, as the Hubble constant $\alpha$ tends to zero. We only consider coordinate transformations that do not mix vector spaces tangent to the sphere with the space of vectors normal to the sphere.
		In other words, we consider transformations between static coordinates $(t,r,\theta,\varphi)$ and coordinates $(T,R,x^{A})$ in the form
		\begin{equation}
			\begin{split}
				T&=T(t,r) \, , \\
				R&=R(t,r) \, , \\
				x^{A}&=x^{A} (\theta, \varphi) \, ,
			\end{split}
			\label{eq:general_conf_transform}
		\end{equation}
		for which the conformally flat metric reads
		\begin{equation}
			g=\Psi\left(T,R\right)\left[-d T^{2}+d R^{2}+R^{2} \sigma_{A B} d x^{2} d x^{3}\right] \, .
			\label{eq:conformallyflatmetric}
		\end{equation}
		
		The transformation form \eqref{eq:general_conf_transform} implies
		\begin{equation}
			\Psi=\frac{r^{2}}{R^{2}} \, .
		\end{equation} 
		We will continue to analyze the problem from a different perspective. The key question is to investigate the freedom of two-dimensional transformation $[\tilde{T}(T,R), \tilde{R}(T,R)]$ which obeys
		\begin{equation}
			\frac{1}{R^{2}}\left(-d T^{2}+d R^{2}\right)=\frac{1}{\tilde{R}^{2}}\left(-d \tilde{T}^{2}+d \tilde{R}^{2}\right) \, .
			\label{eq:CFMSimplify}
		\end{equation}
		We additionally require that the tip
		\begin{equation}
			\{R=0\}=\{\tilde{R}=0\}
		\end{equation}
		remains invariant. Such transformation preserves the form of the metric \eqref{eq:conformallyflatmetric} in obvious way. We proceed with further analysis of \eqref{eq:CFMSimplify} by introducing
		\begin{equation}
			U=T+R \, , \quad V=T-R \, ,
		\end{equation}
		and 
		\begin{equation}
			\tilde{U}=\tilde{T}+\tilde{R} \, , \quad \tilde{V}=\tilde{T}-\tilde{R} \, .
		\end{equation}
		For which \eqref{eq:CFMSimplify} becomes
		\begin{equation}
			\frac{d U d V}{(U-V)^{2}}=\frac{d \tilde{U} d \tilde{V}}{(\tilde{U}-\tilde{V})^{2}} \, .
			\label{eq:CFMSimpuv}
		\end{equation}
		As a next step, we demand that the transformation respects causality i.e. the future pointing vector fields remains future pointing vector fields. It leads to
		\begin{subequations}
			\begin{equation}
				\tilde{U}=F(U) \, ,
			\end{equation}
			\begin{equation}
				\tilde{V}=G(V) \, . 
			\end{equation}
			\label{eq:FG}
		\end{subequations}
		Substituting \eqref{eq:FG} into \eqref{eq:CFMSimpuv}, we find
		\begin{equation}
			\frac{F^{\prime} G^{\prime}}{(F-G)^{2}}=\frac{1}{(U-V)^{2}} \, ,
		\end{equation}
		which further becomes
		\begin{equation}
			\frac{G^{\prime \prime}}{\left(G^{\prime}\right)^{3 / 2}}=\frac{F^{\prime \prime}}{\left(F^{\prime}\right)^{3 / 2}}=\text { const. }
		\end{equation}
		The general solution is given by
		\begin{subequations}
			\begin{equation}
				F(U)=-\frac{1}{A_{\scriptscriptstyle F} U+B_{\scriptscriptstyle F}}+C_{\scriptscriptstyle F} \, ,
			\end{equation}
			\begin{equation}
				G(V)=-\frac{1}{A_{\scriptscriptstyle G} V+B_{\scriptscriptstyle G}}+C_{\scriptscriptstyle G} \, .
			\end{equation}
		\end{subequations}
		Note that, under reasonable assumptions,described near equations \eqref{eq:general_conf_transform}-\eqref{eq:CFMSimplify}, the general solution has a rigid form.

    \bibliography{bibliographyHopfdS}
\end{document}